\documentclass[letterpaper]{article} 
\usepackage{aaai24}  
\usepackage{times}  
\usepackage{helvet}  
\usepackage{courier}  
\usepackage[hyphens]{url}  
\usepackage{graphicx} 
\usepackage{natbib}  
\usepackage{caption} 
\usepackage{algorithm}
\usepackage{algorithmic}

\usepackage{newfloat}
\usepackage{listings}
\DeclareCaptionStyle{ruled}{labelfont=normalfont,labelsep=colon,strut=off} 
\floatstyle{ruled}
\newfloat{listing}{tb}{lst}{}
\floatname{listing}{Listing}
\usepackage{tabularx}
\usepackage{hyperref}
\usepackage{nameref}
\usepackage{xspace}
\usepackage{subcaption}
\usepackage{multirow}
\usepackage{colortbl}
\usepackage{makecell}
\usepackage{array,booktabs}
\usepackage{xcolor}
\usepackage[most]{tcolorbox}
\usepackage{fontawesome}
\usepackage{pgfplots}
\usepackage{pgfplotstable}
\usepackage{fontawesome}
\usepackage{longtable}
\usepackage{enumitem}
\usepackage{calc}
\usepackage{footmisc}[hyperfootnotes=false]
\usepackage{xr}

\pgfplotsset{compat=1.18}

\definecolor{green3}{RGB}{66,179,130}
\definecolor{green2}{RGB}{121,205,169}
\definecolor{green1}{RGB}{196,233,217}

\definecolor{red1}{RGB}{251,219,220}
\definecolor{red2}{RGB}{244,164,166}
\definecolor{red3}{RGB}{236,91,96}
\definecolor{gray1}{RGB}{220,220,220}
\definecolor{cellgray}{rgb}{ .921,  .921, .921}

\definecolor{blue1}{RGB}{101,173,246}
\definecolor{blue2}{RGB}{12,112,212}
\definecolor{blue3}{HTML}{003f5c}

\definecolor{orange1}{RGB}{250,181,97}
\definecolor{orange2}{RGB}{245,138,7}
\definecolor{orange3}{HTML}{7B3F00}

\newcolumntype{P}[1]{>{\raggedright}p{#1}}
\newcolumntype{M}[1]{>{\centering\arraybackslash}m{#1}}
\newcolumntype{L}{>{\RaggedRight}X}

\newcommand*{\belowrulesepcolor}[1]{%
  \noalign{%
    \kern-\belowrulesep 
    \begingroup 
      \color{#1}%
      \hrule height\belowrulesep 
    \endgroup 
  }%
} 
\newcommand*{\aboverulesepcolor}[1]{%
  \noalign{%
    \begingroup 
      \color{#1}%
      \hrule height\aboverulesep 
    \endgroup 
    \kern-\aboverulesep 
  }%
} 

\def\mylegend#1#2{
\resizebox {0.02\linewidth} {6.5pt} {%
\begin{tikzpicture}[]
\begin{axis}[
      axis background/.style={fill=white!30, draw=white!30},
      axis line style={draw=none},
      tick style={draw=none},
      ytick=\empty,
      xtick=\empty,
      ymin=0, ymax=0.70,
      xmin=0, xmax=6]
\addplot [
      ybar interval=.5,
      fill=#2,
      draw=none,
]
	coordinates {(4.5,1) (0,0.30)}; 
\end{axis}%
\end{tikzpicture}%
}%
#1
}

\def\frequencybarchart#1#2#3#4#5#6#7#8{
\resizebox{0.08\linewidth}{7.5pt} {
\begin{tikzpicture}[]
\node[] { \huge \emph{#7}};
\end{tikzpicture}
}
\resizebox {0.81\linewidth} {6.5pt} {%
\begin{tikzpicture}[]
\begin{axis}[
      axis background/.style={fill=gray!30, draw=gray!30},
      axis line style={draw=none},
      tick style={draw=none},
      ytick=\empty,
      xtick=\empty,
      ymin=0, ymax=0.70,
      xmin=0, xmax=6]
\addplot [
      ybar interval=.5,
      fill=green3,
      draw=none,
]
	coordinates {(6*#1,1) (0,0.30)}; 
\addplot [
      ybar interval=.5,
      fill=green2,
      draw=none,
]
	coordinates {(6*(#1+#2),1) (6*#1,1)}; 
\addplot [
      ybar interval=.5,
      fill=gray1,
      draw=none,
]
	coordinates {(6*(#3+#2+#1),1) (6*(#2+#1),1)}; 
\addplot [
      ybar interval=.5,
      fill=red2,
      draw=none,
]
	coordinates {(6*(#4+#3+#2+#1),1) (6*(#3+#2+#1),1)}; 
\addplot [
      ybar interval=.5,
      fill=red3,
      draw=none,
]
	coordinates {(6*(#5+#4+#3+#2+#1),1) (6*(#4+#3+#2+#1),1)}; 
\addplot [
      ybar interval=.5,
      fill=red1,
      draw=none,
]
	coordinates {(6*(#6+#5+#4+#3+#2+#1),1) (6*(#5+#4+#3+#2+#1),1)}; 
\end{axis}%
\end{tikzpicture}%
}
\resizebox{0.08\linewidth}{7.5pt} {
\begin{tikzpicture}[]
\node[] { \huge \emph{#8}};
\end{tikzpicture}
}
}

\def\importancebarchart#1#2#3#4#5#6#7#8#9{
\resizebox{0.08\linewidth}{7.5pt} {
\begin{tikzpicture}[]
\node[] { \huge \emph{#8}};
\end{tikzpicture}
}
\resizebox {0.81\linewidth} {6.5pt} {%
\begin{tikzpicture}[]
\begin{axis}[
      axis background/.style={fill=gray!30, draw=gray!30},
      axis line style={draw=none},
      tick style={draw=none},
      ytick=\empty,
      xtick=\empty,
      ymin=0, ymax=0.70,
      xmin=0, xmax=6]
\addplot [
      ybar interval=.5,
      fill=blue3,
      draw=none,
]
	coordinates {(6*#1,1) (0,0.30)}; 
\addplot [
      ybar interval=.5,
      fill=blue2,
      draw=none,
]
	coordinates {(6*(#1+#2),1) (6*#1,1)}; 
\addplot [
      ybar interval=.5,
      fill=blue1,
      draw=none,
]
	coordinates {(6*(#3+#2+#1),1) (6*(#2+#1),1)}; 

\addplot [
      ybar interval=.5,
      fill=gray1,
      draw=none,
]
	coordinates {(6*(#4+#3+#2+#1),1) (6*(#3+#2+#1),1)}; 
\addplot [
      ybar interval=.5,
      fill=orange1,
      draw=none,
]
	coordinates {(6*(#5+#4+#3+#2+#1),1) (6*(#4+#3+#2+#1),1)}; 
\addplot [
      ybar interval=.5,
      fill=orange2,
      draw=none,
]
	coordinates {(6*(#6+#5+#4+#3+#2+#1),1) (6*(#5+#4+#3+#2+#1),1)}; 
\addplot [
      ybar interval=.5,
      fill=orange3,
      draw=none,
]
	coordinates {(7*(#7+#6+#5+#4+#3+#2+#1),1) (6*(#6+#5+#4+#3+#2+#1),1)}; 
\end{axis}%
\end{tikzpicture}%
}
\resizebox{0.08\linewidth}{7.5pt} {
\begin{tikzpicture}[]
\node[] { \huge \emph{#9}};
\end{tikzpicture}
}
}
\newcommand{\frameworkName}{\textsc{Particip-AI}\xspace}

\title{Particip-AI: A Democratic Surveying Framework for Anticipating Future AI Use Cases, Harms and Benefits}

\author {
    Jimin Mun\textsuperscript{\rm 1},
    Liwei Jiang\textsuperscript{\rm 2, \rm 3},
    Jenny Liang\textsuperscript{\rm 1},
    Inyoung Chung\textsuperscript{\rm 2},
    Nicole DeCario\textsuperscript{\rm 3}, \\
    Yejin Choi\textsuperscript{\rm 2, \rm 3}, 
    Tadayoshi Kohno\textsuperscript{\rm 2},
    Maarten Sap\textsuperscript{\rm 2, \rm 3}
}
\affiliations {
    \textsuperscript{\rm 1}Carnegie Mellon University, Pittsburgh, PA, USA\\
    \textsuperscript{\rm 2}University of Washington, Seattle, WA, USA\\
    \textsuperscript{\rm 3}Allen Institute for AI, Seattle, WA, USA\\
    jmun@andrew.cmu.edu
}

\begin{document}

\maketitle










\begin{abstract}
General purpose AI, such as ChatGPT, seems to have lowered the barriers for the public to use AI and harness its power. 
However, the governance and development of AI still remain in the hands of a few, and the pace of development is accelerating without a comprehensive assessment of risks. 
As a first step towards democratic risk assessment and design of general purpose AI, we introduce \frameworkName, a carefully designed framework for laypeople to speculate and assess AI use cases and their impacts.
Our framework allows us to study more nuanced and detailed public opinions on AI through collecting use cases, surfacing diverse harms through risk assessment under alternate scenarios (i.e., developing and not developing a use case), and illuminating tensions over AI development through making a concluding choice on its development.
To showcase the promise of our framework towards informing democratic AI development, we run a medium-scale study with inputs from 295 demographically diverse participants.
Our analyses show that participants' responses emphasize applications for personal life and society, contrasting with most current AI development's business focus. We also surface diverse set of envisioned harms such as distrust in AI and institutions, complementary to those defined by experts. 
Furthermore, we found that perceived impact of \textit{not} developing use cases significantly predicted participants' judgements of whether AI use cases should be developed, and highlighted lay users' concerns of techno-solutionism.
We conclude with a discussion on how frameworks like \frameworkName can further guide democratic AI development and governance. 
\end{abstract}

\section{Introduction}

In response to rapid adoption of AI and expansion of its application areas, calls for more democratic and comprehensive risk assessment of AI are growing \cite{maslej2023ai,wsjChatGPTsAltman,bengio2023managing}.
Yet, these pose several challenges.
On one hand, the current assessment frameworks and development decisions have largely been guided by experts \cite{solaiman2023evaluating,weidinger2023sociotechnical,barrett2023ai}, overlooking the broadening impact of AI to and its widening usage by everyday, non-expert public \cite{viswanathan-etal-2023-prompt2model,saksPublicAwarenessArtificial2023}. On the other hand, participatory frameworks for AI have been adopted in many specific domains \cite{corbett2023power, friedman2008value,friedman2012envisioning}, but the flexible and instructable nature of general-purpose AI \cite{zoph2022emergent} requires large-scale and diverse participation in \emph{anticipating} use cases of AI in addition to use case development, to comprehensively evaluate its possible impacts. Thus, tackling these challenges is the key towards a more open and less centralized decision-making \cite{widder2023openforbusiness, brynjolfsson2023turing} around AI design and governance.

A necessary first step towards this goal is to build a framework for the non-expert public to share opinions and express critical assessments on AI. Such a framework must be centered around concrete use cases \cite{Trustible2023-vy,AIAct_2023}, since only discussing high-level regulation of general purpose models leads to rules that are too vague to operationalize \cite{Rao2023-hy}. Moreover, the framework should allow the public to consider the alternate reality associated with an AI use case, considering both its development and \textit{non}-development, a contrastive perspective often missing in technology assessment \cite{forlano2023speculative}. Finally, to widely speculate the future of AI, the use cases and development scenarios should cover futuristic, e.g., AI in 5--10 years, use cases and their impact as well.

Towards this goal, we introduce \frameworkName, a framework to gather detailed and nuanced public opinion on AI based on current and future use cases and their impact, inspired by speculative design practices \cite{hohendanner2023exploring,balaram2018artificial} and consequentialist ethics modeling \cite{kohnoEthicalFrameworksComputer2023a, card2020consequentialism}. 
Our framework proposes a four-step process that asks participants to brainstorm possible use cases, imagine and rate their harms and benefits under two alternate scenarios of developing and not developing, and finally, make a choice on the development of the use cases. 

To show the feasibility of our framework, we conduct an online survey with 295 demographically diverse, US-based participants and analyzed their responses, qualitatively and quantitatively, to answer the following research questions.
\begin{enumerate}[label=\bfseries RQ \arabic*,leftmargin=2.7em,labelindent=1em]
    \item[\textbf{RQ1}] To explore overlooked AI development directions and help guide equitable progress through public input, we ask: \textbf{\textit{what current and future use cases of AI are in the public's imagination?}}
    \item[\textbf{RQ2}] As many seemingly beneficial use cases of AI have problems of dual use and failures, we analyze: \textbf{\textit{what are the harms and benefits of the use cases?}}
    \item[\textbf{RQ3}] To gather input beyond technological determinism \cite{littman2022gathering}, we examine: \textbf{\textit{what are the harms and benefits of \underline{not} developing certain applications of AI?}}
    \item[\textbf{RQ4}] To study people's decision processes and conflicting values, we explore: \textbf{\textit{what creates tension between developing and not developing the applications?}}
\end{enumerate}
Our results surface a wide array of anticipated use cases (RQ1), which highlight common themes of interest in improving personal, everyday life, showing diverse interests to augment life through AI and emphasis on the value of AI in making societal impact towards betterment of society as a whole. We find that our participants surface set of harms complementary to taxonomies created by experts (RQ2), for example, raising issues of distrust in institutions and highlighting the need for regulation to protect mental health. Moreover, our findings uncover a set of benefits and harms associated with not developing a use case and highlight a tension in AI's impact on human potential (RQ3). Finally, we find that level of benefits and harms of \textit{not} developing a use case is significantly more correlated with decisions of development, compared to that of developing (RQ4).

To summarize our contribution, we (1) propose a \frameworkName, a novel framework to assess AI use cases and their impact, developed with insights from various field of AI, computer security, and philosophy. We (2) conduct a survey with lay users to showcase \frameworkName's usability. We (3) present results from synthesizing themes such as interest in use cases that emphasize equitable progress through enhancing everyday life and solving societal issues, harm types complementary to the expert-generated, various impacts of not developing a use case, and tensions over value of human work. Finally, we (4) conclude with a discussion on direction of AI development to reflect diverse goals and needs, risks of AI and ways to address regulatory gaps, and tensions over development and techno-solutionism. Our work highlights the promise and importance of including lay publics and diverse voices into the future of AI design and governance.\footnote{See \url{https://github.com/JiMinMun/Particip-AI} for all participant responses.}\footnote{Supplementary material and appendix are at \url{https://arxiv.org/abs/2403.14791}.}
\section{Related Work}


\textbf{Participation in AI.}
While the rapidly growing deployment of AI systems across many sectors has called for meaningful participation~\cite{costanza2018design, delgado2023participatory, queerinai2023queer, pasquale2021if, suresh2022towards}, exploration of such approaches has lagged behind~\cite{birhane2022power, bergman2023representation,durmus2023towards, bender2021dangers}, especially in large-scale AI models~\cite{birhane2022power}. On one hand, previous efforts span ``data labor'' for model optimization such as annotation and feedback~\cite{miceli2022data, bai2022training}, enabling human inputs at granular, instance-level (e.g., human annotation or feedback; \citealp{openaisafety, bai2022training, christiano2023deep}) or at limited stages of AI pipeline (e.g., representative evaluation; \citealp{bergman2023representation}) largely for existing AI systems. On the other hand, many previous works focused on a broad, principles-level, participation including constitutional AI~\cite{bai2022constitutional, collectiveconstitutionalai}, citizen juries~\cite{balaram2018artificial, van2021trading}, public AI policy insights~\cite{adalovelace2023public, NIST2023}, and community collectives~\cite{nekoto2020participatory,queerinai2023queer}. 

Unlike previous works, our work targets the middle-ground by facilitating public assessment of potential real-world AI applications across domains, aligning with recent legislation advocating more application-based approaches~\cite{AIAct_2023}. Our work takes inspiration from design futuring \cite{fry2009design, kozubaev2020expanding}, including speculative design and design fiction, which focuses on non-linear approaches that seeks to challenge and criticize the status quo, explore alternate scenarios, and (re)envision the future \cite{hohendanner2023exploring, farias2022social,baumann2016designing,baumann2018participatory}. Thus, our framework addresses the limitations in previous works through future-looking, use-case oriented questions, which allow for discussion around broader stages of AI pipeline (e.g., usage, design, threats, and opinions of deployment stages) and while we implement the framework as a survey in this work, is not limited to any specific format.

\textbf{Risk Assessment of AI Applications.}
AI applications' far-reaching impact and increasing accessibility among lay users \cite{viswanathan-etal-2023-prompt2model} urges broad, deliberate, and multifaceted assessments of their nuanced and unexpected risks \cite{lubars2019delegability}. 
While many works have developed assessment frameworks of AI risks, most have focused on expert inputs only \cite{solaiman2023evaluating,weidinger2023sociotechnical,barrett2023ai}, neglecting the valuable perspectives of end users impacted by AI-related harms, focusing on broader guidelines (e.g., human rights; \citealp{gabriel2022humrights}), overlooking conflicting values of diverse set of users.

In works that incorporate user inputs for AI risk assessments, there is a noted limitation in accommodating a wide range of diverse and potentially conflicting human values \cite{weidinger2023sociotechnical}. For example, these works cover limited deployment scenarios \cite{buçinca2023aha} or targets specific stage of AI development pipeline and user group (e.g., tool to help AI developers and researchers in prototyping harms; \citealp{wang2024farsight}). To address the challenges of surfacing diverse perspectives and potentially conflicting values, our framework adopts a large-scale participation-based approach with lay-users to anticipate risks associated with current and future AI. 
Moreover, by soliciting lay users' inputs to assess the potential harms and benefits of both \textit{developing} and \textit{not developing} AI applications our work more comprehensively gathers conflicting interests and values. 

\begin{table*}[h]
    \centering
    \small
    \setlength\tabcolsep{1pt}
    \begin{tabularx}{\textwidth}{M{0.015\linewidth}p{0.03\linewidth}p{0.015\linewidth}p{0.1\linewidth}|X}
    \toprule
        \multicolumn{4}{l}{\textbf{Question Numbers}} & \textbf{Content} \\
        \midrule
        \belowrulesepcolor{cellgray}
        \multicolumn{5}{l}{{\cellcolor[rgb]{ .921,  .921, .921}\textbf{Technology Description}}}\\
        \hline
        \raisebox{-0.1\height}\faGear & \multicolumn{2}{l}{Tech-X} & & Imagine an AI technology (let's call it "Tech-X") is developed by tech companies that can follow any instructions to generate new content such as images, human-like language, computer code, etc. To name a few of its capabilities, it can interact with people in a conversational way, write stories, create illustrations and paintings, and answer questions about almost anything. \\
        \hline
        \raisebox{-0.1\height}\faGears & \multicolumn{2}{l}{Tech-X 10} & & Now consider a date five to ten years into the future. Imagine a more sophisticated version of Tech-X (let's call it "Tech-X 10"), which can follow any instruction you give it, has expert-level knowledge or even better, can solve problems creatively, can connect to the internet and other devices, and can process and read massive amounts of data or text within seconds.\\
        \hline
        \multicolumn{5}{l}{{\cellcolor[rgb]{ .921,  .921, .921}\textbf{Sec 1: Use Cases}}}\\
        \hline
        \cellcolor[rgb]{ .95,  .95, .95} & Q1 & \cellcolor[rgb]{ .95,  .95, .95} & Q4 & Do you think a technology like this should exist?\\
        \cellcolor[rgb]{ .95,  .95, .95}\faGear & Q2 & \cellcolor[rgb]{ .95,  .95, .95}\faGears & Q5 & How confident are you in the above answer?\\
        \cellcolor[rgb]{ .95,  .95, .95} & Q3$^{*}$ & \cellcolor[rgb]{ .95,  .95, .95} & Q6$^{*}$ &  Please write some tasks that \{\faGear, \faGears\} could assist or automate in the text box below.\\
        \hline
        \raisebox{-0.1\height}\faSuitcase & Q7$^{*}$ & & & Complete the following sentence by choosing one task from your brainstormed answers that you believe Tech-X or Tech-X 10 will change the most drastically. The task that I think Tech-X / Tech-X 10 would most dramatically change would be in…\\
        \hline
        \multicolumn{5}{l}{{\cellcolor[rgb]{ .921,  .921, .921}\textbf{Sec 2: Harms and Benefits of Developing}}}\\
        \hline
        \cellcolor[rgb]{.95, .95, .95} & Q8$^{*}$ & \cellcolor[rgb]{ .95,  .95, .95} & Q11$^{*}$ \& Q14$^{*}$ & How will Tech-X / Tech-X 10 automating or assisting the task you identified \{\faThumbsOUp, \faThumbsODown\} impact individuals?\\
        \cellcolor[rgb]{ .95,  .95, .95} & Q9$^{*}$ & \cellcolor[rgb]{ .95,  .95, .95} & Q12$^{*}$ \& Q15$^{*}$ & Which groups of people do you think would \{\faThumbsOUp, \faThumbsODown\} the most from the above \{\faThumbsOUp, \faThumbsODown\} impacts?\\
        \multirow{-3}{*}{\cellcolor[rgb]{ .95,  .95, .95}\raisebox{-0.5\height}\faThumbsOUp} & Q10 & \multirow{-3}{*}{\cellcolor[rgb]{ .95,  .95, .95}\raisebox{-0.5\height}\faThumbsODown} & Q13 \& Q16 & How \{\faThumbsOUp, \faThumbsODown\} would it be if Tech-X / Tech-X 10 was used for the following task and had the above \{\faThumbsOUp, \faThumbsODown\} impacts?\\
        \hline
        \multicolumn{5}{l}{{\cellcolor[rgb]{ .921,  .921, .921}\textbf{Sec 3: Harms and Benefits of \textit{Not} Developing}}}\\
        \hline
        \cellcolor[rgb]{ .95,  .95, .95} & Q17$^{*}$ & \cellcolor[rgb]{ .95,  .95, .95} & Q20$^{*}$ &  Now imagine that Tech-X / Tech-X 10 was never used to automate or assist with \{\faSuitcase\}. How will banning or not developing this particular application \{\faThumbsOUp, \faThumbsODown\} impact individuals?\\
        \multirow{-1}{*}{\cellcolor[rgb]{ .95,  .95, .95}\raisebox{-0.5\height}\faThumbsODown} & Q18$^{*}$ & \cellcolor[rgb]{ .95,  .95, .95}\raisebox{-0.5\height}\faThumbsOUp & Q21$^{*}$ & Which groups of people do you think would \{\faThumbsOUp, \faThumbsODown\} the most from the above by banning or not developing this particular application?\\
        \cellcolor[rgb]{ .95,  .95, .95} & Q19 & \cellcolor[rgb]{ .95,  .95, .95} & Q22 & How \{\faThumbsOUp, \faThumbsODown\} would it be if Tech-X / Tech-X 10 was banned or never developed to perform the following task and had the above \{\faThumbsOUp, \faThumbsODown\} impacts?\\
        \hline
        \multicolumn{5}{l}{{\cellcolor[rgb]{ .921,  .921, .921}\textbf{Sec 4: Use Case Opinion}}}\\
        \hline
        \raisebox{-0.2\height}\faPencilSquareO & Q23 & & & After thinking about the benefits and harms of the application and the harms of it not being developed, do you think that this application of the technology should or should not be developed?\\
        & Q24 & & & How confident are you in the above answer?\\
        & Q25 & & & How likely do you think are people to agree that an application of Tech-X / Tech-X 10 that automates or assists with \{\faSuitcase\} \{\faPencilSquareO\}?\\
        \bottomrule
    \end{tabularx}
    \caption{Survey questions in \frameworkName. Questions are summarized due to space constraints. Asterisks (*) denote open-text questions. Within curly brackets are variations such as benefit (\faThumbsOUp) or harm (\faThumbsODown) or input from previous questions, e.g., task (\faSuitcase).}
    \label{tab:survey-qs}
\end{table*}


\section{Methods}
\label{sec:method}


In this section, we first introduce  \frameworkName framework, including the motivation and scope, and the question choices for the survey instrument (\S \ref{ssec:framework-description}). We then describe our methods for collecting and analyzing lay users' inputs (\S \ref{ssec:data-analysis}).


\subsection{\frameworkName Framework}
\label{ssec:framework-description}

\subsubsection{Overview}
\label{ssec:method-survey-overview}
The primary goal of \frameworkName is to effectively elicit the opinions of lay-users on the potential harms and benefits across many near- and far-future AI applications. To create such a framework that addresses pertinent, pivotal questions in AI and to incorporate diverse, bottom-up views, we harness the interdisciplinary expertise of our research team, spanning computer security, public policy, natural language processing (NLP), and AI ethics. We \textit{iteratively} design a survey to reflect our research questions and adopt an \textit{online crowdsourcing platform} for broad and controlled distribution to populations with diverse backgrounds.

To explore how lay users perceive the influences and consequences of \textit{future} AI applications, we prompt users to \textit{imagine} potential use cases of AI and consider speculative harms and benefits of both \textit{developing} and \textit{not developing} such technologies. This fictional inquiry approach \cite{dindler2007fictionalinquiry}, is inspired by various fields, including design fictions \cite{bleekcker2022designfiction} and threat modeling in computer security \cite{evtimovSecurityMachineLearning2020}. In particular, the alternative scenarios (i.e., \textit{to develop} or \textit{not to develop} a use case) involve choosing between two outcomes, reminiscent of hypothetical dilemmas in moral philosophy \cite{bostyn2018dilemma}. 

The option of \textit{developing} an AI use case considers two distinct types of harms: (1) those arising from the \textit{failure or low performance} of AI \cite{raji2022fallacy}, and (2) those resulting from the \textit{malicious misuse} of AI \cite{pohler2024technological}. Finally, acknowledging the distinct real-world impacts of \textit{short-term, near-future}, and \textit{long-term, far-future} AI technologies, \frameworkName guides users to analyze and differentiate the distinct potential harms and benefits presented by AI with varying levels of capabilities.

\subsubsection{Question Design}
\label{ssec:method-survey-questions}
Survey questions are shown in Table~\ref{tab:survey-qs}.

\textbf{Use Cases of AI (RQ1):}
First, participants are asked to imagine the use cases of two variants of AI, \textit{Tech-X} and \textit{Tech-X 10}. 
Tech-X, while fictional, describes a technology similar to current generative AI, i.e., instruction-following with generative output. Tech-X 10, on the other hand, describes a technology five to ten years into the future with a focus on its expert-level knowledge and creative problem-solving. For each variant, participants are asked three questions: whether the technology like the one described should exist or not (Q1, Q4; binary), their confidence in that opinion (Q2, Q5; 5-point Likert), and ideas on use cases of the technology (Q3, Q6; free-text). Finally, participants are asked to choose one brainstormed use case that would be changed \emph{most drastically} through AI (Q7).
\textit{All subsequent questions ask specifically about the use case chosen in this step.}

\textbf{Harms and Benefits of Developing (RQ2):}
Here, participants are asked to anticipate the use case's benefits and the two types of harms (i.e., malicious misuse, failure cases).
Regarding the benefits and each type of harm, participants describe their impact (Q8, Q11, Q14; free-text), the most impacted groups (Q9, Q12, Q15; free-text), and the scale of the impact (Q10, Q13, Q16; 8-point Likert\footnote{\label{fn:anchor}Anchored scale to control for individual user interpretations. See Appendix~\ref{app:survey} for details.}). 


\textbf{Harms and Benefits of NOT Developing (RQ3):}
Next, assuming a hypothetical scenario where the technology is not used for the use case, participants are tasked to describe the impact of potential harms and benefits (Q17, Q20; free-text), most impacted groups (Q18, Q21; free-text), and the scale of impact (Q19, Q22; 8-point Likert\footref{fn:anchor}).

\textbf{Use Case Opinions (RQ4):}
Finally, to understand how participants perceive the permissibility of developing the use case, participants are asked to select whether the application should be developed (Q23; binary), the confidence in that answer (Q24; 5-point Likert), and how likely it would be that others would agree to that opinion (Q25; 8-point Likert\footref{fn:anchor}).



\subsection{Input Collection and Analysis Methodology}
\label{ssec:data-analysis}

We performed targeted recruiting of diverse participants to conduct our survey and a mixture of \textit{quantitative} (for questions with nominal or ordinal scale answers) and \textit{qualitative} analyses (for questions with free-text answers) to extract insights from the collected survey data.

\subsubsection{Participant Recruitment}
\label{ssec:method-survey-recruitment}

We recruited 300 participants on Prolific to conduct the \frameworkName survey.\footnote{https://www.prolific.com/}
To obtain diverse opinions, 
we performed targeted recruiting across five different ethnic groups provided by the platform (i.e., Asian, Black, Mixed, Other, and White),
and across two age groups (i.e., 18 to 48, 49 to 100) that divides the US adult population in approximately half,\footnote{
US Census Data, Accessed on 11/29/2023.} resulting in 10 different groups with 30 participants each. 
These groups were balanced in male and female sex categories.
295 participants responded to the survey.
The survey took median 25.2 minutes. Participants were compensated at the rate of \$9.67/hr. 
Our study was approved by an institutional review board. Limitations due to the platform's recruitment, stratification, and categorization methods are discussed further in \S\ref{sec:limitations}.

Below, we describe a subset of participants' demographics; see Appendix~\ref{app:demographics} for a full description.
While we recruited equal number of demographically stratified participants, imbalances occurred due to the specification of self-identified racial categories and lack of responses in certain groups.
Participants were White (29.8\%), Black (21.7\%), Asian (20.0\%), Other (10.8\%), and Mixed (11.9\%). 
Participants largely identified as men (47.8\%) and women (47.1\%).

\subsubsection{Qualitative Analysis}
We apply qualitative analysis to free-form answers, using concept coding~\cite{saldana2021coding} practices augmented with GPT-4\footnote{\texttt{gpt-4-1106-preview}} to aid in applying the human-generated codebook on the large-scale survey dataset.
These questions include AI use cases (Q3, Q6, Q7), impact of benefits (Q8, Q20), impact of harms (Q11, Q14, Q17), and most impacted groups (Q9, Q12, Q15, Q18, Q21). See Appendix~\ref{app:codes} for the codes and their definition.

\paragraph{Open Coding with Human Annotators}
With three authors from our research team, we developed codebooks by performing open coding on approximately 25\% of the data (80 samples). For questions on tasks and impact of harms and benefits (Q7, Q8, Q11, Q14, Q17, Q20), three authors developed codebooks independently, and then merged them into a shared codebook upon discussion with unanimous agreement. Next, the three authors applied the merged codebooks to all questions of each user sample. Individual low-level codes are grouped into a high-level \textit{theme} during data analysis. In addition, we pre-process brainstormed tasks (Q3, Q6) using GPT-4 to standardize their expressions (see Appendix~\ref{app:parsing} for further details). Finally, given the brevity and directness of answers, questions pertaining to groups impacted (Q9, Q12, Q15, Q18, Q21) were coded by a single author. See Appendix~\ref{app:coding} for further details.

\paragraph{Closed Coding with GPT-4 Augmentation}
Manually coding responses is prohibitive when the sample size is large; frontier language models such as GPT-3 and GPT-4 have shown promise in automatic qualitative coding \cite{xiao2023supporting,matter2024close}. Thus, we applied GPT-4 to perform closed coding of the remaining 75\% of the samples using the codebooks developed by our research team during the open coding process.
We use the first two samples from the held-out data that we manually coded as few-shot examples to prompt GPT-4, and we evaluate human to GPT-4 agreement using the remaining 78 samples.
In line with previous research \cite{xiao2023supporting}, GPT-4 has moderate (0.41-0.60) to substantial (0.61-0.80) agreement \cite{mchugh2012interrater} with the human annotators.\footnote{Inter-rater agreement scores based on Cohen's $\kappa$~\cite{cohen1960coefficient} on validation set for each question: (Q7; $\kappa$=.59, Q8; $\kappa$=.51, Q11; $\kappa$=.66, Q14; $\kappa$=.67, Q17; $\kappa$=.62, Q20; $\kappa$=.58, Q9; $\kappa$=.77, Q12; $\kappa$=.85, Q15; $\kappa$=.87, Q18; $\kappa$=.85, Q21; $\kappa$=.82)} See Appendix~\ref{app:closed-coding} for detailed prompts, settings, and additional metrics.

\subsubsection{Quantitative Analysis.}
Quantitative analysis is conducted on data in nominal or ordinal scales, including opinions on whether a use case should be developed (Q1, Q4, Q23; binary), the scale of impact (Q10, Q13, Q16, Q19, Q22; Likert) and participants' rating of their confidence and anticipated agreement (Q24, Q25; Likert). We aggregate the percentage of responses for opinions, take the mean by theme for the harms and benefits scales, and conduct an exploratory factor analysis for the effects of harms and benefits on opinions of use case development.\footnote{For effects analysis, we converted participants' answers numerically: opinion ($-1$=``Should not be developed'', $1$=``Should be developed''), confidence ($-4$=``Should not be developed'' $\times$ ``Extremely confident'', $4$=``Should be developed''$\times$``Extremely confident'') and perceived agreement ($-8$=``Should not be developed'' $\times$ ``Highly likely'', $8$=``Should be developed'' $\times$ ``Highly likely'')}

\section{Results}
\label{sec:results}

In this section, we present and discuss the results of our survey based on \frameworkName framework with 295 participants. Before delving into the specific use cases, participants showed an overall positive attitude towards the general descriptions of the AI technology with 86.1\% (Q1) and 85.4\% (Q4) responding that Tech-X and Tech-X 10 ``Should exist''.


\begin{table*}[h]
    \centering
    \small
    \setlength\tabcolsep{1pt}
    \begin{tabularx}{\textwidth}{p{0.23\linewidth}|X|p{0.052\linewidth}|p{0.045\linewidth}|p{0.045\linewidth}}
    \toprule
        & & \multicolumn{3}{l}{\textbf{\hfill \% responses}} \\
        \textbf{\texttt{Code} / \textsc{Theme}} & \textbf{Quote} & \hfil \textbf{Q3} & \hfil \textbf{Q6} & \hfil \textbf{Q7} \\
        & & \hfil$1032\times$ & \hfil$992\times$ & \hfil$295\times$ \\
        \hline
        \multicolumn{2}{l|}{\cellcolor[rgb]{ .921,  .921, .921}\textbf{\textsc{Domain}}} & \cellcolor[rgb]{ .921,  .921, .921} \hfil\textbf{55.0} & \cellcolor[rgb]{ .921,  .921, .921} \hfil\textbf{63.2} & \cellcolor[rgb]{ .921,  .921, .921} \hfil\textbf{67.8} \\
        \hline
        \texttt{Artistic expression} & \textit{``Make abstract art with my dog's picture''} (P13, Q3) & \hfil 13.4 & \hfil 3.4 & \hfil 5.4 \\
        \texttt{Medical} & \textit{``Automating medical research''} (P54, Q6) & \hfil 4.9 & \hfil 10.5 & \hfil 13.9 \\
        \texttt{Education} & \textit{``Teaching me how to code in different languages''} (P104, Q7) & \hfil 11.1 & \hfil 9.7 & \hfil 10.5  \\
        \texttt{Research} & \textit{``Revolutionize scientific research''} (P203, Q6) & \hfil 3.0 & \hfil 8.5 & \hfil 10.5  \\
        \texttt{Translation} & \textit{``Translations while traveling''} (P292, Q3) & \hfil 8.0 & \hfil 1.3 & \hfil 4.1  \\
        \hline
        \multicolumn{2}{l|}{\cellcolor[rgb]{ .921,  .921, .921}\textbf{\textsc{Support type}}} & \cellcolor[rgb]{ .921,  .921, .921} \hfil\textbf{39.0} & \cellcolor[rgb]{ .921,  .921, .921} \hfil\textbf{39.3} & \cellcolor[rgb]{ .921,  .921, .921} \hfil\textbf{40.7} \\
        \hline
        \texttt{Efficient data analysis} & \hangindent=1em \textit{``Assist in personalized medicine by analyzing genetic data, medical histories, and current research...''} (P242, Q6) & \hfil 8.9 & \hfil 17.4 & \hfil 19.0  \\
        \texttt{Professional consulting service} & \hangindent=1em \emph{``Mental health diagnosis and intervention...since...professionals often gets overwhelmed with their work."} (P203, Q7)  & \hfil 2.1 & \hfil 6.3 & \hfil 7.8 \\
        \texttt{Writing assistance} & \emph{``Adapting resumes to different job postings''} (P146, Q3)  & \hfil 11.5 & \hfil 3.6 & \hfil 2.7 \\
        \hline
        \multicolumn{2}{l|}{\cellcolor[rgb]{ .921,  .921, .921}\textbf{\textsc{Personal life}}}  & \cellcolor[rgb]{ .921,  .921, .921} \hfil\textbf{45.7} & \cellcolor[rgb]{ .921,  .921, .921} \hfil\textbf{39.2} & \cellcolor[rgb]{ .921,  .921, .921} \hfil\textbf{29.5} \\
        \hline
        \texttt{Everyday task automation} & \textit{``Summarizing important email content into a list''} (P293, Q3)  & \hfil 25.4 & \hfil 22.1 & \hfil 14.2 \\
        \texttt{Everyday life assistance} & \hangindent=1em \textit{``Assisting me with planning meal ideas that meet my family's dietary needs...[to] take much weight off my mental plate.''} (P249, Q7)  & \hfil 17.8 & \hfil 15.6 & \hfil 13.9 \\
        \hline
        \multicolumn{2}{l|}{\cellcolor[rgb]{ .921,  .921, .921}\textbf{\textsc{Goal of the use case}}} & \cellcolor[rgb]{ .921,  .921, .921} \hfil\textbf{26.5} & \cellcolor[rgb]{ .921,  .921, .921} \hfil\textbf{25.9} & \cellcolor[rgb]{ .921,  .921, .921} \hfil\textbf{33.6} \\
        \hline
        \texttt{Personal life productivity} & \textit{``Assisting with managing time spent on activities"} (P100, Q3)  & \hfil 8.8 & \hfil 6.0 & \hfil 10.2 \\
        \texttt{Creativity} & \hangindent=1em \textit{``Creating unique entertainment options that cater to individuals and evolve with them over time''} (P265, Q6) & \hfil 12.4 & \hfil 4.6 & \hfil 6.4 \\
        \hline
        \multicolumn{2}{l|}{\cellcolor[rgb]{ .921,  .921, .921}\textbf{\textsc{Society}}} & \cellcolor[rgb]{ .921,  .921, .921} \hfil\textbf{0.3} & \cellcolor[rgb]{ .921,  .921, .921} \hfil\textbf{8.7} & \cellcolor[rgb]{ .921,  .921, .921} \hfil\textbf{9.5} \\
        \hline
        \texttt{Societal issues} & \hangindent=1em \textit{``Predictive models that improve health and environment challenges.''} (P28, Q7) & \hfil0.3 & \hfil 8.7 & \hfil 9.5  \\
        \hline
       \multicolumn{2}{l|}{\cellcolor[rgb]{ .921,  .921, .921}\textbf{\textsc{Work}}} & \cellcolor[rgb]{ .921,  .921, .921} \hfil\textbf{5.2} & \cellcolor[rgb]{ .921,  .921, .921} \hfil\textbf{6.8} & \cellcolor[rgb]{ .921,  .921, .921} \hfil\textbf{10.8} \\
        \hline
        \texttt{Human labor replacement} & \textit{``Replacing humans in customer interaction jobs''} (P282, Q6) & \hfil 0.8 & \hfil 4.8 & \hfil 8.1  \\
        \texttt{Workplace productivity} & \textit{``Generate job reports that would take human hours''} (P86, Q6) & \hfil 4.5 & \hfil 2.5 & \hfil 4.1  \\
        \hline
        \multicolumn{2}{l|}{\cellcolor[rgb]{ .921,  .921, .921}\textbf{\textsc{Other}}} & \cellcolor[rgb]{ .921,  .921, .921} \hfil\textbf{0.8} & \cellcolor[rgb]{ .921,  .921, .921} \hfil\textbf{2.2} & \cellcolor[rgb]{ .921,  .921, .921} \hfil\textbf{4.4} \\
        \hline
        \texttt{New code} & \textit{``Find a way for the AI to destroy its own AI self''} (P25, Q6) & \hfil 0.8 & \hfil 2.0 & \hfil 3.7  \\
        \bottomrule
    \end{tabularx}
    \caption{Tasks (Q3, Q6, Q7): percentage of occurrence for \textsc{theme} and top few most frequent \texttt{codes} with representative quotes.}
    \label{tab:q3-q6-q7-theme}
\end{table*}

\subsection{RQ1. Current and Future Use Cases of AI}

To answer this question, we analyze the brainstormed use cases for current (Q3) and future technology (Q6), as well as the tasks that would be most drastically impacted by AI (Q7). We grouped the codes (see Appendix~\ref{app:use-case}) into the high-level \textit{themes} to assist analyzing the results and highlight general trends (see Table~\ref{tab:q3-q6-q7-theme}): \textsc{domain}, \textsc{support type}, realms of impact (i.e., \textsc{work}, \textsc{personal life}, and \textsc{society}), and \textsc{goal of the use case}.

\subsubsection{Current and Future AI Use Cases}
Participants brainstormed a similar number of use cases across themes for both current (Q3, Tech-X; avg. of 3.5) and future (Q6, Tech-X 10; avg. of 3.4 tasks) technology.
Across both questions, participants most commonly mentioned the \textsc{domain} of use cases (55.0\%; Q3, 63.2\%; Q6) compared to \textsc{support type}, \textsc{goal}, or realms of impact.
However, use cases differed in their distributions within the theme: notably, those for future technology emphasized domains such as \texttt{medical} (10.5\%), \texttt{education} (9.7\%), and \texttt{research} (3.0\%) whereas those for the current version discussed \texttt{artistic expression} (13.4\%), \texttt{education} (11.1\%), and \texttt{translation} (8.0\%). 
\textsc{Personal life} applications occured more frequently for Tech-X (45.7\%) compared to Tech-X 10 (39.2\%).
In contrast, tasks surrounding impact to \textsc{society} grew most drastically from Tech-X (0.3\%) to Tech-X 10 (8.7\%), suggesting people's interest in future AI applications to address societal issues.
See Table~\ref{tab:q3-q6-q7-theme} for detailed distribution of the themes and codes.

\subsubsection{Participant Selected Use Cases}

Among tasks described as most revolutionized by AI, \textsc{domain} of applications was the most common theme (67.8\%), covering \texttt{medical} (13.9\%), \texttt{education} (10.5\%), and \texttt{research} (10.5\%) domains. The second most prevalent theme was \textsc{support type} (40.7\%) containing top use cases related to \texttt{efficient data analysis} (19.0\%) and \texttt{professional consulting service} (7.8 \%).
As in previous questions, \textsc{personal life} (29.5\%) related tasks were discussed more frequently compared to \textsc{work} (10.8\%) and \textsc{society} (9.5\%) (see Appendix~\ref{app:ssec-use-case-personal-life}). Notably, participants selected more use cases that impact \textit{the society} compared to previous questions (see Appendix~\ref{app:ssec-use-case-society}).

\subsection{RQ2. Harms and Benefits of the Use Cases}
To answer this queston, we analyzed participants' anticipated harms and benefits of their use case (qualitative; Q8, Q11, Q14), groups that could be harmed or benefited the most (qualitative; Q9, Q12, Q15), and the scale of impact (quantitative; Q10, Q13, Q16). Analysis on questions such as anticipated groups affected by developing use case (Q9, Q12, Q15) can be found in Appendix~\ref{app:ssec-groups-dev}.

\begin{table*}[h]
    \centering
    \small
    \setlength\tabcolsep{1pt}
    \begin{tabularx}{\textwidth}{p{0.22\linewidth}|X|p{0.045\linewidth}|p{0.045\linewidth}}
    \toprule
        & & \multicolumn{2}{c}{\textbf{\% responses}} \\
        \textbf{Code} & \textbf{Quote \& Scale of Impact (Q13 / Q16)} & \hfil \textbf{Q11} & \hfil \textbf{Q14} \\
        & & \hfil $295\times$ & \hfil $295\times$ \\
        \hline
        \multicolumn{2}{l|}{\cellcolor[rgb]{ .921,  .921, .921}\rlap{\makebox[0.65\linewidth][r]{\textbf{5.82$\pm$1.94 / 5.29$\pm$2.05}}}\textbf{\textsc{Social \& psychological effect}}} & \cellcolor[rgb]{ .921,  .921, .921} \hfil \textbf{35.3} & \cellcolor[rgb]{ .921,  .921, .921} \hfil \textbf{26.4} \\
        \hline
        \texttt{Manipulate people} &  \textit{``People would lose control potentially over important data, ideas...''} (P113, Q11) & \hfil 12.9 & \hfil 1.0 \\
        \texttt{Misinformation} & \hangindent=1em \textit{``It could give false information and confuse people as to where they don’t know which source of information to trust"} (P126, Q11) & \hfil 12.5 & \hfil 10.8 \\
        \texttt{Mental harm} & \hangindent=1em \textit{``Loss of confidence and motivation: Repeated misunderstandings and failed interactions could lead to frustration and a reluctance to engage in...learning.''} (P280, Q14) & \hfil 12.2 & \hfil 9.5 \\
        \texttt{Social isolation} & \textit{``It would harm...relationships...maybe [leading] to ostracism or loss of trust.''} (P200, Q14) & \hfil 2.4 & \hfil 4.4 \\
        \hline
        \multicolumn{2}{l|}{\cellcolor[rgb]{ .921,  .921, .921}\rlap{\makebox[0.65\linewidth][r]{\textbf{5.39$\pm$1.70 / 4.63$\pm$1.75}}} \textbf{\textsc{Economic impact}}} & \cellcolor[rgb]{ .921,  .921, .921} \hfil \textbf{32.5} & \cellcolor[rgb]{ .921,  .921, .921} \hfil \textbf{33.6} \\
        \hline
        \texttt{Financial disturbance} &  \textit{``People would lose jobs and incomes''} (P93, Q11) & \hfil 16.3 & \hfil 20.7 \\
        \texttt{Economic disturbance} & \textit{``Shortage in suppliers or a raise in costs"} (P126, Q11) & \hfil 12.9 & \hfil 9.8 \\
        \texttt{Waste resources or time} & \textit{``Potentially leading to misguided decisions [and] wasted resources...''} (P250, Q14) & \hfil 1.0 & \hfil 9.5 \\
        \hline
        \multicolumn{2}{l|}{\cellcolor[rgb]{ .921,  .921, .921}\rlap{\makebox[0.65\linewidth][r]{\textbf{6.32$\pm$1.57 / 6.10$\pm$1.60}}}\textbf{\textsc{Safety \& security risk}}} & \cellcolor[rgb]{ .921,  .921, .921} \hfil \textbf{21.7} & \cellcolor[rgb]{ .921,  .921, .921} \hfil \textbf{7.8} \\
        \hline
        \texttt{Data security \& privacy risk} & \textit{``It could be...compromising users' private data''} (P235, Q14) & \hfil 10.5 & \hfil 3.1 \\
        \texttt{Extinction} & \textit{``human race eliminated by machines''} (P220, Q11) & \hfil 5.8 & \hfil 2.0 \\
        \texttt{Aid criminal} & \textit{``It will lead to theft''} (P275, Q14) & \hfil 4.1 & \hfil 2.7 \\
        \hline
        \multicolumn{2}{l|}{\cellcolor[rgb]{ .921,  .921, .921}\rlap{\makebox[0.65\linewidth][r]{\textbf{6.70$\pm$1.40 / 6.28$\pm$1.59}}} \textbf{\textsc{Physical effect}}} & \cellcolor[rgb]{ .921,  .921, .921} \hfil \textbf{16.9} & \cellcolor[rgb]{ .921,  .921, .921} \hfil \textbf{29.8} \\
        \hline
        \texttt{Physical harm} &  \textit{``It could lead to serious injury  or death.''} (P215, Q14) & \hfil 12.9 & \hfil 23.7 \\
        \texttt{Negative health \& well-being} & \textit{``People would become more unhealthy''} (P175, Q11) & \hfil 3.4 & \hfil 8.5 \\
        \hline
        \multicolumn{2}{l|}{\cellcolor[rgb]{ .921,  .921, .921}\rlap{\makebox[0.65\linewidth][r]{\textbf{5.16$\pm$1.76 / 5.43$\pm$2.12}}}\textbf{\textsc{Quality \& reliability issues of AI}}} & \cellcolor[rgb]{ .921,  .921, .921} \hfil \textbf{15.9} & \cellcolor[rgb]{ .921,  .921, .921} \hfil \textbf{24.1} \\
        \hline
        \texttt{Incorrect AI output} &\textit{``Providing incorrect or incomplete medical diagnostics''} (P221, Q14) & \hfil 9.8 & \hfil 13.6 \\
        \texttt{Distrust AI} & \textit{``People would lose trust in technology''} (P72, Q11) & \hfil 5.1 & \hfil 5.8 \\
        \hline
        \multicolumn{2}{l|}{\cellcolor[rgb]{ .921,  .921, .921}\rlap{\makebox[0.65\linewidth][r]{\textbf{4.41$\pm$1.80 / 4.42$\pm$2.10}}}\textbf{\textsc{Impeding human development \& learning}}} & \cellcolor[rgb]{ .921,  .921, .921} \hfil \textbf{12.9} & \cellcolor[rgb]{ .921,  .921, .921} \hfil \textbf{12.5} \\
        \hline
        \texttt{Overreliance} & \textit{``[People] will not learn to do anything on their own''} (P274, Q14) & \hfil 9.5 & \hfil 6.8 \\
        \texttt{Impede learning} &\textit{``Diminished capacity for original ideas, maybe even critical thinking''} (P221, Q11) & \hfil 3.7 & \hfil 3.7 \\
        \texttt{Hinder career} & \textit{``The reputation of the developers would be ruined''} (P125, Q14) & \hfil 1.0 & \hfil 4.1 \\
        \hline
        \multicolumn{2}{l|}{\cellcolor[rgb]{ .921,  .921, .921}\rlap{\makebox[0.65\linewidth][r]{\textbf{5.70$\pm$2.01 / 5.48$\pm$2.09}}}\textbf{\textsc{Reducing quality \& reliability of society}}} & \cellcolor[rgb]{ .921,  .921, .921} \hfil \textbf{12.2} & \cellcolor[rgb]{ .921,  .921, .921} \hfil \textbf{9.2} \\
        \hline
        \texttt{Distrust institution} & \textit{``The negative impact would be...decreased trust in the medical professionals...''} (P62, Q14) & \hfil 5.8 & \hfil 5.4 \\
        \texttt{Legal issues} & \textit{``Increased lawsuits.''} (P95, Q14) & \hfil 3.1 & \hfil 4.1 \\
        \hline
        \multicolumn{2}{l|}{\cellcolor[rgb]{ .921,  .921, .921}\rlap{\makebox[0.65\linewidth][r]{\textbf{5.95$\pm$2.25 / 5.97$\pm$1.94}}}\textbf{\textsc{General harm}}} & \cellcolor[rgb]{ .921,  .921, .921} \hfil \textbf{6.8} & \cellcolor[rgb]{ .921,  .921, .921} \hfil \textbf{11.5} \\
        \hline
        \texttt{General harm} & \textit{``More people would be hurt''} (P105, Q11) & \hfil 4.7 & \hfil 10.2 \\
        \texttt{Range} & \textit{``Could cause anything from minor issues to loss of life''} (P271, Q14) & \hfil 2.0 & \hfil 1.7 \\
        \hline
        \multicolumn{2}{l|}{\cellcolor[rgb]{ .921,  .921, .921}\rlap{\makebox[0.65\linewidth][r]{\textbf{7.50$\pm$0.84 / 6.57$\pm$1.45}}}\textbf{\textsc{Reducing progress}}} & \cellcolor[rgb]{ .921,  .921, .921} \hfil \textbf{2.0} & \cellcolor[rgb]{ .921,  .921, .921} \hfil \textbf{4.7} \\
        \hline
        \texttt{Environmental harm} & \textit{``It could negatively impact fighting climate change''} (P23, Q14) & \hfil 1.4 & \hfil 1.7 \\
        \texttt{Hinder science} &  \textit{``Delay scientific advancement and progress''} (P276, Q14) & \hfil 0.7 & \hfil 3.1 \\
        \hline
        \multicolumn{2}{l|}{\cellcolor[rgb]{ .921,  .921, .921}\rlap{\makebox[0.65\linewidth][r]{\textbf{3.30$\pm$3.22 / 4.28$\pm$2.83}}}\textbf{\textsc{Other}}} &  \cellcolor[rgb]{ .921,  .921, .921} \hfil \textbf{4.4} & \cellcolor[rgb]{ .921,  .921, .921} \hfil \textbf{7.1} \\
        \hline
        \texttt{N/A} & \textit{``Many things''} (P216, Q11) & \hfil 2.4 & \hfil 2.7 \\
        \texttt{No harm} & \textit{``I can't think of any [harms]''} (P242, Q11) & \hfil 1.7 & \hfil 3.7 \\
        \bottomrule
    \end{tabularx}
    \caption{Harms of developing (Q11, Q14): percentage of occurrence for \textsc{theme} with scale of impact (Q13, Q16) and corresponding top few most frequent \texttt{codes} with representative quotes.}
    \label{tab:q11-theme}
\end{table*}

\subsubsection{Harms of Developing}
\label{sssec:harms-dev}
The harms of developing the selected use cases were grouped into ten high-level themes (see Table \ref{tab:q11-theme}; left). 
We analyzed harms due to misuses (Q11) and poor performance (Q14) separately. 
For harms due to \textit{misuses or unintended consequences}, participants most often mentioned \textsc{social and psychological effect} (35.3\%) followed by \textsc{economic impact} (32.5\%). Within \textsc{social and psychological effect}, the most common concerns were \texttt{manipulation of people} (12.9\%) (e.g., ``\textit{control and manipulate information for human exploitation}'' by P114), \texttt{misinformation} (12.5\%), and \texttt{mental harm} (12.2\%).
For harms caused by the poor performance of technology, participants most commonly discussed \textsc{economic impact} (33.6\%) at the personal and societal level, such as \texttt{financial disturbance} (20.7\%) and \texttt{economic disturbance} (9.8\%).
The second most discussed harm due to failure cases was \textsc{physical effect} (29.8\%), such as \texttt{physical harm} (23.7\%) and \texttt{negative impact to health and well being} (8.5\%).

While the two types of harm showed different distributions of themes, their scale of impact was similar. Among all themes, \textsc{reducing progress} (7.50$\pm$0.84; Q13, 6.57$\pm$1.45; Q16) and \textsc{physical} (6.70$\pm$1.40; Q13, 6.28$\pm$1.59; Q16) had the biggest scale of impact (see Table~\ref{tab:q11-theme}). 
While \textsc{economic impact} was a frequent theme overall, its perceived impact was lower (5.39$\pm$1.70; Q13, 4.63$\pm$1.75; Q16), especially for poor performance harms. 


\begin{table*}[h]
    \centering
    \small
    \setlength\tabcolsep{1pt}
    \begin{tabularx}{\textwidth}{p{0.22\linewidth}|X|p{0.095\linewidth}}
    \toprule
        & & \hfil \textbf{\% responses} \\
        \textbf{Code} & \textbf{Quote \& Scale of Impact (Q10)} & \hfil \textbf{Q8} \\
        & & \hfil $295\times$ \\
        \hline
        \multicolumn{2}{l|}{\cellcolor[rgb]{ .921,  .921, .921}\rlap{\makebox[0.6\linewidth][r]{\textbf{5.13$\pm$1.86}}}\textbf{\textsc{Reinvest human capital}}} & \cellcolor[rgb]{ .921,  .921, .921} \hfil \textbf{52.5} \\
        \hline
        \texttt{Personal life efficiency} &  \textit{``Save time, effort, and energy...[and] allow a layperson to accomplish this task.''} (P19) & \hfil 35.6 \\
        \texttt{Personal growth} &  \hangindent=1em \textit{``Since it's data driven, individual performances will be vigorously assessed and suggest ways by which an individual can improve.''} (P47) & \hfil 16.9 \\
        \texttt{Reduce mundane work} & \hangindent=1em \textit{``I would be able to focus on relationships and team building versus menial manager tasks that AI could complete for me.''} (P157) & \hfil 13.2 \\
        \hline
        \multicolumn{2}{l|}{\cellcolor[rgb]{ .921,  .921, .921}\rlap{\makebox[0.6\linewidth][r]{\textbf{5.38$\pm$1.75}}}\textbf{\textsc{Economic gain}}} & \cellcolor[rgb]{ .921,  .921, .921} \hfil \textbf{43.7} \\
        \hline
        \texttt{General efficiency} & \hangindent=1em \textit{``Companies will not need to have as many employees...because they'll be able to automate much of the workload...which will increase company profits.''} (P209) & \hfil 31.5 \\
        \texttt{Financial gain} & \textit{``It will save cost of different diagnostic tests.''} (P211) & \hfil 17.6 \\
        \hline
        \multicolumn{2}{l|}{\cellcolor[rgb]{ .921,  .921, .921}\rlap{\makebox[0.6\linewidth][r]{\textbf{5.58$\pm$1.62}}}\textbf{\textsc{Resource accessibility}}} & \cellcolor[rgb]{ .921,  .921, .921} \hfil \textbf{35.9} \\
        \hline
        \texttt{Information accessibility} & \hangindent=1em \textit{``People need quick and reliable answers because not a lot of people have time for themselves...[and] can't deeply engage in topics they encounter in daily life.''} (P53) & \hfil 18.0 \\
        \texttt{Resource accessibility} & \textit{``It would give people more equity and assistance.''} (P99) & \hfil 15.6 \\
        \hline
        \multicolumn{2}{l|}{\cellcolor[rgb]{ .921,  .921, .921}\rlap{\makebox[0.6\linewidth][r]{\textbf{6.96$\pm$1.25}}}\textbf{\textsc{Improve societal issues}}} & \cellcolor[rgb]{ .921,  .921, .921} \hfil \textbf{31.9} \\
        \hline
        \texttt{Improve medical care} &  \textit{``Health care would be cheaper (hopefully) and more accessible to everyone''} (P109) & \hfil 13.2 \\
        \texttt{Scientific research innovation} &  \textit{``Research would be able to be done at a faster pace.''} (P159) & \hfil 11.2 \\
        \hline
        \multicolumn{2}{l|}{\cellcolor[rgb]{ .921,  .921, .921}\rlap{\makebox[0.6\linewidth][r]{ \textbf{6.11$\pm$1.57}}}\textbf{\textsc{Reduce error}}} & \cellcolor[rgb]{ .921,  .921, .921} \hfil \textbf{16.9} \\
        \hline
        \texttt{Less human error} & \textit{``It could remove certain human biases''} (P171) & \hfil 10.8 \\
        \texttt{Information quality} & \textit{``It could quickly detect lies said by politicians.''} (P229) & \hfil 9.8 \\
        \hline
        \multicolumn{2}{l|}{\cellcolor[rgb]{ .921,  .921, .921}\rlap{\makebox[0.6\linewidth][r]{\textbf{6.50$\pm$1.50}}}\textbf{\textsc{Improve quality of personal life}}} & \cellcolor[rgb]{ .921,  .921, .921} \hfil \textbf{15.3} \\
        \hline
        \texttt{Improve well-being \& health} &  \textit{``My family would have a healthier diet \& they would live better lives.''} (P238) & \hfil 10.2 \\
        \texttt{Improve mental health} & \textit{``It would reduce the cases of mental illness in lonely people.''} (P96) & \hfil 5.4 \\
        \hline
        \multicolumn{2}{l|}{\cellcolor[rgb]{ .921,  .921, .921}\rlap{\makebox[0.6\linewidth][r]{\textbf{6.76$\pm$1.17}}}\textbf{\textsc{Improve quality of social life}}} & \cellcolor[rgb]{ .921,  .921, .921} \hfil \textbf{8.1} \\
        \hline
        \texttt{Better communication} & \textit{``People would be able to communicate in different languages in real-time.''} (P176) & \hfil 5.4 \\
        \texttt{Social interaction} & \hangindent=1em \textit{``It would help me navigate through various social situations and problems, thus improving my social life.''} (P200) & \hfil 2.0 \\
        \hline
        \multicolumn{2}{l|}{\cellcolor[rgb]{ .921,  .921, .921}\rlap{\makebox[0.6\linewidth][r]{\textbf{5.25$\pm$2.17}}}\textbf{\textsc{Other}}} & \cellcolor[rgb]{ .921,  .921, .921} \hfil \textbf{5.4} \\
        \hline
        \texttt{New code} &  \textit{``Help others with problems''} (P232) & \hfil 2.7 \\
        \bottomrule
    \end{tabularx}
    \caption{Benefits of developing (Q8): Percentage of occurrence for \textsc{theme} with scale of impact (Q10) and corresponding top few most frequent \texttt{codes} with representative quotes.}
    \label{tab:q8-theme}
\end{table*}

\subsubsection{Benefits of Developing}
\label{sssec:benefits-dev}

The benefits of selected use cases are grouped into eight themes (see Table \ref{tab:q8-theme}; right). The most prominent theme was \textsc{reinvesting human capital} (52.5\%), within which, \texttt{personal life efficiency} (35.6\%) to ``\textit{save time effort and energy}'' (P19) in personal life was mentioned the most followed by \texttt{personal growth} (16.9\%), and \texttt{reducing mundane work} (13.2\%).
The second most frequent theme was \textsc{economic gain} (43.7\%), such as \texttt{general efficiency} (31.5\%) and \texttt{financial gain} (17.6\%). 

While \textsc{reinvesting human capital} was the most frequently observed benefit, its scale of impact (5.13$\pm$1.86) was lower compared to \textsc{improving quality of social life} (6.76$\pm$1.17) and \textsc{improving quality of personal life} (6.50$\pm$1.50). 
This suggests that while AI offers efficiency in reinvesting human capital, the more influential positive impact comes from improving the quality of life.

\subsection{RQ3. Harms and Benefits of \textit{Not} Developing Certain Applications of AI}
We analyzed participants' answers on harms and benefits of not developing (qualitative; Q17, Q20), groups that could be harmed or benefited the most (qualitative; Q18, Q21), and the scale of impact (quantitative; Q19, Q22) to answer this question. See Appendix~\ref{app:ssec-groups-nd} for analysis on questions about anticipated groups affected by use case (Q18, Q20).

\begin{table*}[h]
    \centering
    \small
    \setlength\tabcolsep{1pt}
    \begin{tabularx}{\textwidth}{p{0.182\linewidth}|X|p{0.095\linewidth}}
    \toprule
        & & \hfil \textbf{\% responses} \\
        \textbf{Code} & \textbf{Quote \& Scale of Impact (Q19)} & \hfil \textbf{Q17} \\
        & & \hfil $295\times$ \\
        \hline
        \multicolumn{2}{l|}{\cellcolor[rgb]{ .921,  .921, .921}\rlap{\makebox[0.6\linewidth][r]{\textbf{3.56$\pm$2.12}}}\textbf{\textsc{Limiting human potential}}} & \cellcolor[rgb]{ .921,  .921, .921} \hfil \textbf{32.9} \\
        \hline
        \texttt{Waste resources or time} &  \textit{``I waste so much time on these types of activities. Time that could be spent on productive things...''} (P39) & \hfil 13.2 \\
        \texttt{Inefficiency} & \hangindent=1em \textit{``Government and public agencies will continue to operate in a wasteful and ineffective manner.''} (P76) & \hfil 12.2 \\
        \texttt{Impede personal growth} & \textit{``People would not be able to reach their potentials.''} (P106) & \hfil 11.9 \\
        \hline
        \multicolumn{2}{l|}{\cellcolor[rgb]{ .921,  .921, .921}\rlap{\makebox[0.6\linewidth][r]{\textbf{4.14$\pm$2.04}}}\textbf{\textsc{Lose information \& accessibility to resources}}} & \cellcolor[rgb]{ .921,  .921, .921} \hfil \textbf{26.1} \\
        \hline
        \texttt{Lose assistance} & \textit{``Immigrants would not recieve [sic] translation support easily.''} (P122) & \hfil 11.5 \\
        \hangindent=1em \texttt{Lose solution or service} & \textit{``Homeless need easier more accessible help...''} (P206) & \hfil 9.8 \\
        \hline
        \multicolumn{2}{l|}{\cellcolor[rgb]{ .921,  .921, .921}\rlap{\makebox[0.6\linewidth][r]{\textbf{2.73$\pm$2.59}}}\textbf{\textsc{Other}}} & \cellcolor[rgb]{ .921,  .921, .921} \hfil \textbf{23.4} \\
        \hline
        \texttt{No harm} & \textit{``It wouldnt [sic] necessarily be harmful''} (P216) & \hfil 16.6 \\
        \texttt{New code} & \textit{``It may be an emergency''} (P1) & \hfil 4.1 \\
        \hline
        \multicolumn{2}{l|}{\cellcolor[rgb]{ .921,  .921, .921}\rlap{\makebox[0.6\linewidth][r]{\textbf{4.64$\pm$2.08}}}\textbf{\textsc{Less innovation}}} &  \cellcolor[rgb]{ .921,  .921, .921} \hfil \textbf{15.9} \\
        \hline
        \texttt{Delay in innovation} & \textit{``It could slow down progress against climate change''} (P23) & \hfil 9.5 \\
        \texttt{Less innovation} & \textit{``New technology would not be used to help man kind.''} (P87) & \hfil 8.5 \\
        \hline
        \multicolumn{2}{l|}{\cellcolor[rgb]{ .921,  .921, .921}\rlap{\makebox[0.6\linewidth][r]{\textbf{4.07$\pm$2.40}}}\textbf{\textsc{Social \& psychological effect}}} & \cellcolor[rgb]{ .921,  .921, .921} \hfil \textbf{15.6.5} \\
        \hline
        \texttt{Stress \& overworked} & \textit{``It would increase the workload and time spent on tedious tasks.''} (P181) & \hfil 10.8 \\
        \texttt{Mental harm} &  \textit{``It would deprive people of an opportunity to address their loneliness''} (P96) & \hfil 4.7 \\
        \hline
        \multicolumn{2}{l|}{\cellcolor[rgb]{ .921,  .921, .921}\rlap{\makebox[0.6\linewidth][r]{\textbf{4.78$\pm$2.19}}}\textbf{\textsc{Less progress in solving societal issues}}} & \cellcolor[rgb]{ .921,  .921, .921} \hfil \textbf{13.9.5} \\
        \hline
        \texttt{Hinder medical care} & \textit{``Many individuals will continue suffering from ailments that...worsen in time.''} (P117) & \hfil 8.5 \\
        \texttt{Misinformation} &  \textit{``Some people...find bad answers on the internet that make things worse''} (P246) & \hfil 3.7 \\
        \hline
        \multicolumn{2}{l|}{\cellcolor[rgb]{ .921,  .921, .921}\rlap{\makebox[0.6\linewidth][r]{\textbf{3.70$\pm$2.27}}}\textbf{\textsc{Economic \& business impact}}} & \cellcolor[rgb]{ .921,  .921, .921} \hfil \textbf{11.5.5} \\
        \hline
        \texttt{Financial disturbance} &  \textit{``Individuals might lack access to highly personalized and good retirement strategies''} (P292) & \hfil 4.7 \\
        \texttt{Economic disturbance} & \textit{``Increases cost and reduce employment''} (P16) & \hfil 4.1 \\
        \hline
        \multicolumn{2}{l|}{\cellcolor[rgb]{ .921,  .921, .921}\rlap{\makebox[0.6\linewidth][r]{\textbf{4.39$\pm$1.64}}}\textbf{\textsc{Limited to human capabilities}}} &  \cellcolor[rgb]{ .921,  .921, .921} \hfil \textbf{9.5} \\
        \hline
        \texttt{Human error} &  \textit{``Humans are biased...and often unable to combine various fields of thought.''} (P88) & \hfil 6.8 \\
        \texttt{Hinder creative work} & \textit{``It could hinder some people's ability to create.''} (P120) & \hfil 2.7 \\
        \hline
        \multicolumn{2}{l|}{\cellcolor[rgb]{ .921,  .921, .921}\rlap{\makebox[0.6\linewidth][r]{\textbf{5.14$\pm$2.62}}}\textbf{\textsc{Physical effect}}} & \cellcolor[rgb]{ .921,  .921, .921} \hfil \textbf{8.5} \\
        \hline
        \texttt{Physical harm} & \textit{``it could've saved a lot of lives''} (P152) & \hfil 5.8 \\
        \texttt{Health issues} & \textit{``My health will suffer.''} (P30) & \hfil 3.7 \\
        \bottomrule
    \end{tabularx}
    \caption{Harms of \textit{not} developing (Q17): percentage of occurrence for \textsc{theme} with scale of impact (Q19) and corresponding top few most frequent \texttt{codes} with representative quotes.}
    \label{tab:q17-theme}
\end{table*}

\subsubsection{Harms of Not Developing}
Responses on harms of not developing the use case are grouped into nine high level themes (see Table~\ref{tab:q17-theme}).
The most common themes were \textsc{limiting human potential} (32.9\%) and \textsc{lose information and accessibility to resources} (26.1\%).
By not developing the application, it's anticipated that there will be more \texttt{wasted resources or time} (13.2\%) or \texttt{inefficiency} (12.2\%), e.g., ``\textit{where people's lives are being wasted on unfulfilling labor for low pay}'' (P110).
Another major concern involves \texttt{losing assistance} for the task (11.5\%) and \texttt{losing accessibility to solution and service} (9.8\%).
Unlike the harms of developing (\S\ref{sssec:harms-dev}), 23.4\% of answers were categorized as \textsc{other}, within which many answers mentioned there being \texttt{no harm} (16.6\%), indicating that harms of not developing often does not exist or is harder to imagine compared to harms of developing (e.g., ``\textit{if it never gets develop [sic] we won't know what we are missing out on}'').

Regarding the scale of impact (Q19; see Table~\ref{tab:q17-theme}), \textsc{physical effect} had the highest perceived impact of harm (5.14$\pm$2.62), similar to the scale of impact indicated in harms of developing. 
Participants also anticipate a high impact of \textsc{less progress in solving societal issues} (4.78$\pm$2.19), which, considering previous results, conveys solving societal issues an important beneficial area of AI. 

\begin{table*}[h]
    \centering
    \small
    \setlength\tabcolsep{1pt}
    \begin{tabularx}{\textwidth}{p{0.24\linewidth}|X|p{0.095\linewidth}}
    \toprule
        & & \hfil \textbf{\% responses} \\
        \textbf{Code} & \textbf{Quote \& Scale of Impact (Q22)} & \hfil \textbf{Q20} \\
        & & \hfil $295\times$ \\
        \hline
        \multicolumn{2}{l|}{\cellcolor[rgb]{ .921,  .921, .921}\rlap{\makebox[0.6\linewidth][r]{\textbf{3.95$\pm$2.08}}}\textbf{\textsc{Human growth \& potential}}} & \cellcolor[rgb]{ .921,  .921, .921} \hfil \textbf{43.4} \\
        \hline
        \texttt{Less dependent on technology} & \hangindent=1em \textit{``People would think critically and rely on the thoughts of other human beings who have a more nuanced understanding of real life situations then AI ever could.''} (P113) & \hfil 26.4 \\
        \texttt{Learning skills \& knowledge} &  \textit{``It would make it so more people would strive to learn the local language''} (P122) & \hfil 20.7 \\
        \texttt{Human interaction dependence} &  \textit{``I might have to communicate that I need help and hopefully would bring us together.''} (P249) & \hfil 13.2 \\
        \hline
        \multicolumn{2}{l|}{\cellcolor[rgb]{ .921,  .921, .921}\rlap{\makebox[0.6\linewidth][r]{\textbf{3.54$\pm$2.23}}}\textbf{\textsc{Economic impact / security}}} & \cellcolor[rgb]{ .921,  .921, .921} \hfil \textbf{22.4} \\
        \hline
        \texttt{Job security} &  \textit{``It will not take over peoples' jobs.''} (P251) & \hfil 17.6 \\
        \texttt{Financial benefit} & \hangindent=1em \textit{``The insurance companies and doctors...make more money off of multiple visits''} (P272) & \hfil 6.8 \\
        \hline
        \multicolumn{2}{l|}{\cellcolor[rgb]{ .921,  .921, .921}\rlap{\makebox[0.6\linewidth][r]{\textbf{3.05$\pm$2.79}}}\textbf{\textsc{Other}}} & \cellcolor[rgb]{ .921,  .921, .921} \hfil \textbf{18.6} \\
        \hline
        \texttt{No benefit} &  \textit{``I don't see any benefits''} (P272) & \hfil 10.8 \\
        \texttt{New code} &  \textit{``The company could put in more effort''} (P148) & \hfil 4.4 \\
        \hline
        \multicolumn{2}{l|}{\cellcolor[rgb]{ .921,  .921, .921}\rlap{\makebox[0.6\linewidth][r]{\textbf{3.66$\pm$2.41}}}\textbf{\textsc{Less bad AI usage}}} & \cellcolor[rgb]{ .921,  .921, .921} \hfil \textbf{15.6} \\
        \hline
        \texttt{Less improper or unethical use} &  \textit{``It would not allow for the potential harmful uses of the ai assistance''} (P235) & \hfil 9.5 \\
        \texttt{More privacy} &  \textit{``It would protect information of all and keep breaches at a minimum''} (P293) & \hfil 4.4 \\
        \hline
        \multicolumn{2}{l|}{\cellcolor[rgb]{ .921,  .921, .921}\rlap{\makebox[0.6\linewidth][r]{\textbf{3.34$\pm$2.06}}}\textbf{\textsc{Increase transparency, control, \& reliability}}} & \cellcolor[rgb]{ .921,  .921, .921} \hfil \textbf{12.5} \\
        \hline
        \texttt{More attentive} & \textit{``It could foster more personal involvement...in one’s investment choices''} (P31) & \hfil 7.1 \\
        \texttt{Human control} & \hangindent=1em \textit{``It allows for more deliberate, controlled, and transparent progress...fostering public trust and the responsible development of technology.''} (P204) & \hfil 4.1 \\
        \hline
        \multicolumn{2}{l|}{\cellcolor[rgb]{ .921,  .921, .921}\rlap{\makebox[0.6\linewidth][r]{\textbf{3.26$\pm$2.33}}}\textbf{\textsc{No changes}} \cellcolor[rgb]{ .921,  .921, .921}} & \cellcolor[rgb]{ .921,  .921, .921} \hfil \textbf{7.8} \\
        \hline
        \texttt{Maintain status quo} & \textit{``There would really be no change in society, it would remain the same''} (P216) & \hfil 5.8 \\
        \texttt{Other non-AI solutions} & \textit{``Advances in medicine woud still occur with use of other technologies and methods''} (P45) & \hfil 2.0 \\
        \hline
        \multicolumn{2}{l|}{\cellcolor[rgb]{ .921,  .921, .921}\rlap{\makebox[0.6\linewidth][r]{\textbf{4.75$\pm$2.71}}}\textbf{\textsc{Beneficial side effects of not using AI}}}  & \cellcolor[rgb]{ .921,  .921, .921} \hfil \textbf{2.7} \\
        \hline
        \texttt{Better health} & \textit{``People...[would]...seek qualified medical assistance, which could save their life.''} (P43) & \hfil 2.0 \\
        \texttt{Environmental} & \textit{``AI requires a lot of energy so not developing it will be good for the environment''} (P209) & \hfil 0.7 \\
        \bottomrule
    \end{tabularx}
    \caption{Benefits of \textit{not} developing (Q20): percentage of occurrence for \textsc{theme} with scale of impact (Q22) and corresponding top few most frequent \texttt{codes} with representative quotes.}
    \label{tab:q20-theme}
\end{table*}

\subsubsection{Benefits of Not Developing}
Benefits of not developing the use cases had seven high level themes (see Table~\ref{tab:q20-theme}).
The most reported benefit of not developing AI was \textsc{human growth and potential} (43.4\%), such as \texttt{less dependence on tech} (26.4\%), \texttt{learning skills and knowledge} (20.7\%), and \texttt{increased human interaction and dependence} on one another (13.2\%).
The second most common benefit was \textsc{economic impact and economic security}, such as \texttt{job security} (17.6\%) and \texttt{financial benefits} (6.8\%).
Regarding the scale of benefit (Q22, see Table~\ref{tab:q20-theme}), \textsc{beneficial side effects of not using AI} (e.g., \texttt{better health} and \texttt{environmental impact}) had the highest impact (4.75$\pm$2.71).
\textsc{human growth and potential} also had a high impact (3.95$\pm$2.71), showing an tension in delaying technological progress for the sake of not \textsc{limiting human potential}.



\begin{table*}[h!]
\begin{center}
\small
\begin{tabular}{l r r r r r r}
\hline
 & \multicolumn{2}{c}{Development Opinion (Q23)} & \multicolumn{2}{c}{Confidence (Q24)} & \multicolumn{2}{c}{Agreement (Q25)} \\
 & Coefficient (SE) & $p$-value & Coefficient (SE) & $p$-value & Coefficient (SE) & $p$-value \\
\midrule
Benefits of Developing (Q10) & $0.16$ $(0.07)$ & $<.05$ & $0.17$ $(0.06)$ & $<.01$ & $0.20$ $(0.06)$ & $<.01$ \\
Harms of Developing (Q13) & $-0.08$ $(0.07)$ & $0.30$ & $-0.12$ $(0.06)$ & $<.05$ &  $-0.10$ $(0.07)$ & $0.19$\\
Harms of Developing (Q15) & $-0.12$ $(0.07)$ & $0.08$ & $-0.18$ $(0.07)$ & $<.01$ & $-0.08$ $(0.07)$ & $0.22$\\
Harms of Not Developing (Q19)                      & $\mathbf{0.27}$ $(0.07)$ & $<.001$ & $\mathbf{0.41}$ $(0.06)$ & $<.001$ & $\mathbf{0.29}$ $(0.06)$ & $<.001$\\
Benefits of Not Developing (Q22)                   & $\mathbf{-0.20}$ $(0.06)$ & $<.001$ & $\mathbf{-0.23}$ $(0.05)$ & $<.001$ & $\mathbf{-0.22}$ $(0.06)$ & $<.001$\\
\bottomrule
\end{tabular}
\caption{Effects of each benefit and harms scale to the development opinion \{-1, 1\}, confidence \{-4, 4\}, and agreement \{-8, 8\}. All scales were normalized, and negative values denote opinions ``Should not be developed''. Standard error is in parenthesis.}
\label{table:decision-scale-lmer}
\end{center}
\end{table*}

\subsection{RQ4. Tension between Developing and Not Developing the Applications}
\label{ssec:tension}

%
We analyzed participants' opinions on use case development, their decision confidence, and perceived agreement from others to identify the source of tension.
Most participants answered that the use case ``should'' (83\%) rather than ``should not'' (17\%) be developed. 

\subsubsection{Factors in Tensions over Development}
\label{sssec:tension-factors}
To examine how considering harms and benefits impacted opinions on whether a use case should be developed, we ran a linear mixed effects model with the opinion as dependent variable and the scale of benefits and harms as the independent variables (see Table~\ref{table:decision-scale-lmer}). 
Surprisingly, \textit{Harms of not developing} consistently showed the most significant effect on opinions ($\beta=0.27$, $p<0.001$), confidence ($\beta=0.41$, $p<0.001$), and agreement ($\beta=0.29$, $p<0.001$) that the application \textit{should} be developed.
Similarly, \textit{benefits of not developing} showed the most significant effect on the opinion that the application \textit{should not} be developed (opinions; $\beta=-0.20$, $p<0.001$, confidence; $\beta=-0.23$, $p<0.001$, agreement; $\beta=-0.22$, $p<0.001$).
These results highlight that considering \textit{not developing} scenarios provides deeper insights into people's opinions about AI development than \textit{developing} scenarios alone.
Interestingly, despite the harms and benefits of not developing are use case specific, these answers generally reflected participants' general attitudes toward AI. Specifically, among people who believe their selected use cases should not be developed, 54.9\% and 52.9\% of them also think Tech-X (Q1) and Tech-X 10 (Q4) ``Should not exist,'' respectively, higher than proportions from all the participants (13.9\%; Q1, 14.6\%; Q4).

\subsubsection{Case Analysis}
\label{sssec:tension-case-analysis}


Ordering domains and themes by the number of use cases participants said should not be developed, applications that impact \textsc{work} (28.1\%) ranked first, followed by \textsc{society} (21.4\%) and \textsc{goal} (18.2\%); see Appendix~\ref{app:ssec-usecase-should-not} for further analysis. 
Examples of similar tasks that participants said should and should not be developed are shown in Table~\ref{tab:tension-examples} (see Appendix~\ref{app:use-case}).
P189 (``Should not be developed'') and P32 (``Should be developed'') both discuss reduced accessibility in the harms of not developing an AI assisted transportation system. However, they focus on different benefits of not developing where P32 states that those negatively affected by not developing ``\textit{will miss out on being more independent}'' whereas P189 reflects on ``\textit{human workers}'' keeping their jobs. 
Meanwhile, P175 (``Should not be developed'') and P10 (``Should be developed'') discuss similar lack of assistance in addressing mundane task of buying groceries, but P10 noticed less reliance whereas P175 noticed not developing would increase human interaction by forcing people to ``\textit{go outside and interact in society}''. 
Interestingly, we find that in answers to benefits of not developing, participants who thought the application should not be developed focused more on alternate solutions having additional benefits not addressable by technology whereas those who thought the application should be developed focused on the absence of possible harms from technology.

\section{Discussion \& Conclusion}
\label{sec:discussion-conclusion}
To address the need for lay user participation in anticipating the harms and benefits of AI use cases, we introduced \frameworkName, a framework to collect diverse AI use cases and examine the benefits and harms of both developing and \textit{not} developing them. 
Using our framework, we collected AI use cases from nearly 300 demographically diverse participants.
We now discuss the implications of our findings on future work towards democratic AI development and policy.

\textbf{Benefits of AI: Augmenting Life and Social Good.} 
The brainstorming exercises in our framework uncovered a new array of AI usage highlighting personal life applications to augment everyday life, and societal applications to enrich the lives of everyone. 
At a personal level, participants expressed interest in automating daily tasks, assisting personal growth including mental and physical health, and better allocation of resources (\S\ref{sssec:benefits-dev}), echoing the need for AI design to allow greater stakeholder liberties \cite{bondiEnvisioningCommunitiesParticipatory2021}. 

Participants showed strong interest in using AI to solve significant societal problems from advances in medicine to addressing inequality, global warming, and world hunger. These use cases, present a stark contrast to the current directions of AI development geared towards work and business productivity \cite{maslej2023ai}. Thus, future studies should explore methods to satisfy these public needs like digital commons \cite{verdegem2022commons}, moving beyond profitability.

\textbf{Harms Envisioned by Lay Users.} Our framework also enabled participants to reason through harms and benefits of AI use cases, where a unique set of harms emerged. Our results show that lay users can anticipate the impacts of AI in their daily lives, complementary to technical experts' assessment \cite{solaiman2023evaluating,weidinger2023sociotechnical}. While the themes had some overlap with \citet{solaiman2023evaluating}, additional or more detailed harms were uncovered such as distrust in AI and technology, distrust in institutions, and stalled progress (\S\ref{sssec:harms-dev}). 
Experts also discussed harms that were not as common in participant responses such as environmental costs, and data labor, highlighting the complementary value of the approaches. We project that AI literacy could further empower non-expert public to surface and discuss more diverse and relevant harms \cite{long2020literacy}.

\textit{Psychological harms} such as manipulation, misinformation, and mental harm were among the most common concerns (\S\ref{sssec:harms-dev}), however, have been largely overlooked in current regulatory and academic discussions on AI. Some emerging works have examined the psychological impact of AI and automation (e.g., depression \cite{vidal2020social,mcclure2022case}, influence on autonomy \cite{hackenburg2023evaluating,jakesch2023co}, over-reliance \cite{ma2023understanding}); but the negative impacts of AI remains largely under-explored~\cite{farahany2023battle}. Concerningly, these intangible yet impactful harms would not be effectively remedied through law and policy like EU AI Act \cite{palka2023ai} or US liability case law \cite{cheong2023us}. Therefore, further studies to understand how AI affects mental health are paramount to establishing frameworks that can reveal harms to hold different actors accountable. 
\textbf{Techno-solutionism and Tensions of (Not) Developing.}
As seen with many examples \cite{haven2020philanthropy}, AI applied without careful consideration can exacerbate the existing inequalities by creating a hierarchy of the technology owner and the recipient, especially through its opaqueness \cite{mirca2021nonhuman}. In addition to the harms of AI such as disparate performance on majority vs. minority groups \cite{Buolamwini2018genderShades,sap-etal-2019-risk,scheuerman2019gender}, risk of dual-use \cite{kaffee-etal-2023-thorny}, and imposition of norms \cite{santyliang2023nlpositionality}, the foregone benefits of non-technical solutions such as job creation, human involvement, and community building should be further studied and considered when discussing risks and harms of AI, as illustrated by our framework. 

By collecting and analyzing harms and benefits under two alternate scenarios (to develop and not develop), we aimed to understand the reasoning behind users' decisions. Our qualitative analyses showed that participants often emphasized the benefits of non-technical solutions, such as increased social interaction, job security, and positive impacts toward indirect stakeholders, when they opted for not developing the use case (\S\ref{sssec:tension-case-analysis}). This highlights the need for discussions of often overlooked non-technical solutions and their benefits to various stakeholders, particularly to those vulnerable and marginalized, beyond the default persona of technology (i.e., a culturally prototypical user, often straight white tech-savvy men) \cite{sim2023marginalized}. Thus, anticipation of consequences \cite{Do2023-kh} in making development decisions could be a promising direction towards an inclusive progress.

Participants' responses on harms and benefits of not developing the AI system also highlighted tensions around human growth and potential. Not developing a use case could reduce the efficiency of allocating human resources, but the absence of AI applications could fortify human worth and independence, spurring investment in human knowledge and skills. This dilemma underscores the tension between human's value in creation activities and its perceived competition with that of the machines. This resonates with creator groups' call for protective regulations for their work \cite{authorsguild,authorsguildArtificialIntelligence} and researchers' warning against greater inequality from AI-induced productivity \cite{brynjolfsson2023turing,littman2022gathering,moradi2020future,cheong2023us}. Given these concerns, 
researchers, developers, and companies should consider immediate and long term impacts of AI in labor to maintain the value of human work. In developing AI, a focus on implementing participatory approaches to ensure positive and mitigate negative impacts on affected communities \cite{bondiEnvisioningCommunitiesParticipatory2021,delgado2023participatorydesign,floridi2021design}. Additionally, regulatory measures and economic policies must aim to ensure human value and equality in the distribution of AI-generated benefits \cite{littman2022gathering}.
\section*{Ethical Considerations}
\label{sec:limitations}
Our research endeavor, while aimed at inclusivity, predominantly involved participants from the United States who are English speakers, a demographic feature that carries significant ethical implications regarding the generalizability and inclusivity of our findings \cite{leeWhoIncludedHuman2021}. We recognize the potential marginalization of non-English speakers and individuals outside of the U.S., and the ethical responsibility to ensure that the frameworks we investigate are adaptable across diverse cultural and linguistic contexts. This calls for future research to extend our framework to a broader array of participants, including those from varied cultural backgrounds and those less familiar with technology. Doing so would not only enhance the robustness of our findings but also uphold the ethical principle of inclusivity.

Moreover, the use of an online survey platform, such as Prolific, inherently skews our sample towards individuals who are more comfortable with and have access to technology. This presents an ethical consideration regarding the digital divide and the potential exclusion of technologically underserved populations. Ethical research practice necessitates actively seeking ways to include these populations to avoid reinforcing existing disparities.

Additionally, while we manually verified the quality of our study's data, crowdsourcing-based studies are decentralized and difficult to guarantee the reliability of the data. Moreover, in our analysis, we used GPT-4 to code our data; therefore, with our released data, future work could explore other approaches such as clustering, fully manual coding, coding based on predefined taxonomy, or improving GPT-4's coding abilities. The survey wording, formatting and ordering could have affected the participant answers~\cite{ordereffect1981}, and future works should further explore effects of ordering and phrasing of the questions and descriptions of the future technologies. 

\section*{Acknowledgements}
We thank the Prolific workers for their thoughtful answers. 
This work was supported, in whole or in part, by the Block Center for Technology and Society at Carnegie Mellon University, in part by DARPA under the ITM program (FA8650-23-C-7316), and by the University of Washington Tech Policy Lab, which receives support from the William and Flora Hewlett Foundation, the John D. and Catherine T. MacArthur Foundation, Microsoft, and the Pierre and Pamela Omidyar Fund at the Silicon Valley Community Foundation.
Additionally, Jenny T. Liang was supported by the National Science Foundation under grants DGE1745016 and DGE2140739.

\bibliography{custom}
\clearpage
\appendix
\section{Survey}
\label{app:survey}
\subsection{Scale Anchors}
We anchored the scales we used for harms, benefits, and perceived likelihood of agreement. 
For example, we used the following scale to anchor the benefits: ``Slightly beneficial (comparable to a free meal)'', ``Somewhat beneficial'' (comparable to improving public transportation), ``Very beneficial (comparable to saving a life)'', and ``Extremely 
beneficial (comparable to stopping a war or curing a disease)''.
For the scale of harms, we used anchors such as comparable to jaywalking, theft, arson, and terrorism in order of increasing degree of harm, and for agreement we used no alignment, slight preferences, equally split, majority winner but an ongoing debate, and a clear winner without further debate, as our anchors in increasing agreement.

\begin{table*}[!htpb]
    \footnotesize
    \centering
    \begin{tabular}{P{0.2\linewidth}P{0.055\linewidth}|P{0.04\linewidth}p{0.055\linewidth}|P{0.15\linewidth}p{0.065\linewidth}|P{0.15\linewidth}p{0.065\linewidth}}
    \toprule
         \textbf{Racial Identity} & \textbf{\textit(N) (\%)} & \textbf{Age} & \textbf{\textit{N} (\%)} & \textbf{Gender Identity} & \textbf{\textit{N} (\%)} & \textbf{Education} & \textbf{\textit{N} (\%)} \\
         \midrule
        White or Caucasian & 88 (29.8) & 18-24 & 26 (8.8) & Woman & 141 (47.8) & Bachelor’s degree & 119 (40.3)\\
        Black or African American & 64 (21.7) & 25-34 & 60 (20.3) & Man & 139 (47.1) & Graduate degree$^{*}$ & 55 (18.6)\\
        Asian & 59 (20.0) & 35-44 & 49 (16.6) & Prefer not to disclose & 4 (1.4) & Some college $^{*}$ & 48 (16.3)\\
        Other & 32 (10.8) & 45-54 & 85 (28.8) & Genderqueer$^{*}$ & 4 (1.4) & Associates degree$^{*}$ & 36 (12.2)\\
        Prefer not to say & 10 (3.4) & 55-64 & 56 (19.0) & Additional identity$^{*}$ & 4 (1.36) & High school diploma$^{*}$ & 32 (10.8)\\
        Pacific Islander$^{*}$ & 2 (0.7) & 65+ & 19 (6.4) & Multiple Identities & 2 (0.7) & Prefer not to say & 3 (1.0)\\
        Native American$^{*}$ & 5 (1.7) &  &  & Agender & 1 (0.3) & Some high school$^{*}$ & 2 (0.71)\\
        Mixed & 35 (11.9) &  &  & &\\
    \bottomrule
    \end{tabular}
    \caption{Racial, age, gender identities and education level of participants. Asterisk (*) denotes labels shortened due to space. Additionally, ``Other'' racial identities included Hispanic/Latinx (\textit{N}=26), and ``Additional identity'' included Non-binary (\textit{N}=4). See Appendix~\ref{app:demographics} for more detail.}
    \label{tab:demographics}
\end{table*}
\begin{table*}[htpb]
    \centering
    \footnotesize
    \begin{tabular}{ll|ll|ll|ll}
    \toprule
         Race - Other & \textit{N} & Transgender & \textit{N} & Sexuality & \textit{N} & Political Leaning & \textit{N}\\
         \hline
Hispanic/Latinx & 25 & No & 281 & Straight (heterosexual) & 216 & Liberal & 96\\
multi, Asian, caucasian & 1 & Yes & 8 & Bisexual & 39 & Moderate & 74\\
Arabic Middle Eastern & 1 & Prefer not to disclose & 6 & Gay & 11 & Strongly liberal & 66\\
Brown & 1 &  &  & Lesbian & 7 & Conservative & 40\\
Middle Eastern & 1 &  &  & Pansexual & 6 & Strongly conservative & 11\\
Black and white & 1 &  &  & Prefer not to disclose & 6 & Prefer not to say & 8\\
West Indian & 1 &  &  & Asexual & 3 &  & \\
Indigenous American & 1 &  &  & Other & 2 &  & \\
Mexican American & 1 &  &  & More than one applicable & 5 &  & \\
Caribbean & 1 &  &  &  &  &  & \\
multi & 1 &  &  &  &  &  & \\
Sephardic jew & 1 &  &  &  &  &  & \\
asian, white and middle eastern & 1 &  &  &  &  &  & \\
Hebrew & 1 &  &  &  &  &  & \\
Cajun & 1 &  &  &  &  &  & \\
\bottomrule
    \end{tabular}
    \caption{Additional demographic identities}
    \label{tab:app-dem1}
\end{table*}
\begin{table*}[htpb]
    \centering
    \footnotesize
    \begin{tabularx}{\textwidth}{Xl|Xl|Xl|Xl}
    \toprule
        Longest Residence & \textit{N} & Employment & \textit{N} & Occupation (Top 10) & \textit{N} & Religion & \textit{N} \\
    \hline
United States of America & 288 & Employed, 40+ & 132 & Other: please specify & 29 & Christian & 98\\
Philippines & 3 & Employed, 1-39 & 66 & Health Care and Social Assistance & 26 & Agnostic & 49\\
Guyana & 1 & Not employed, looking for work & 40 & Professional, Scientific, and Technical Services & 26 & Catholic & 43\\
Nigeria & 1 & Retired & 18 & Prefer not to answer & 24 & Nothing in particular & 41\\
China & 1 & Other: please specify & 13 & Retail Trade & 24 & Atheist & 36\\
Cuba & 1 & Not employed, NOT looking for work & 13 & Arts, Entertainment, and Recreation & 22 & Something else & 10\\
&  & Disabled, not able to work & 10 & Information & 22 & Buddhist & 6\\
&  & Prefer not to disclose & 3 & Educational Services & 21 & Jewish & 6\\
&  &  &  & Finance and Insurance & 20 & Muslim & 6\\
&  &  &  & Construction & 13 &  & \\
 \bottomrule
    \end{tabularx}
    \caption{Additional demographics}
    \label{tab:app-dem2}
\end{table*}
\subsection{Participant Demographics}
\label{app:demographics}
The main demographics of participants are included in Table~\ref{tab:demographics}. 
Additional demographics collected are shown in Table~\ref{tab:app-dem1} and \ref{tab:app-dem2}.

\begin{table*}
    \centering
    \footnotesize
    \begin{tabular}{p{0.42\linewidth}p{0.11\linewidth}|p{0.40\linewidth}}
    \toprule
    \textbf{AI Literacy Question} & \textbf{Theme} & \textbf{Distribution} \\
    \midrule
    Q1. I can identify the AI technology employed in the applications and products I use. & Awareness & \importancebarchart{0.037}{0.058}{0.088}{0.166}{0.251}{0.336}{0.064}{3.7\%}{6.4\%}\\
    Q2. I can skillfully use AI applications or products to help me with my daily work. & Usage  & \importancebarchart{0.027}{0.054}{0.078}{0.166}{0.292}{0.254}{0.129}{2.7\%}{12.9\%}\\
    Q3. I can choose the most appropriate AI application or product from a variety for a particular task. & Evaluation & \importancebarchart{0.037}{0.034}{0.112}{0.214}{0.271}{0.234}{0.098}{3.7\%}{9.8\%}\\
    Q4. I always comply with ethical principles when using AI applications or products. & Ethics & \importancebarchart{0.010}{0.014}{0.027}{0.166}{0.159}{0.336}{0.288}{1.0\%}{28.8\%}\\
    Q5$^R$. I am never alert to privacy and information security issues when using AI applications or products. & Ethics & \importancebarchart{0.136}{0.217}{0.207}{0.251}{0.112}{0.054}{0.024}{13.6\%}{2.4\%}\\
    Q6. I am always alert to the abuse of AI technology. & Ethics & \importancebarchart{0.027}{0.054}{0.064}{0.217}{0.275}{0.254}{0.108}{2.7\%}{10.8\%}\\
    \midrule
    \multicolumn{3}{c}{\mylegend{Strongly disagree}{blue3} \mylegend{Disagree}{blue2} \mylegend{Somewhat disagree}{blue1} \mylegend{Neutral}{gray1} \mylegend{Somewhat agree}{orange1} \mylegend{Agree}{orange2} \mylegend{Strongly agree}{orange3}} \\
    \midrule
    Q7. How frequently do you use generative AI (i.e., artificial intelligence that is capable of producing high quality texts, images, etc. in response to prompts) products such as ChatGPT, Bard, DALL·E 2, Claude, etc.? & Usage & \importancebarchart{0.102}{0.227}{0.146}{0}{0.258}{0.166}{0.101}{10.2\%}{10.1\%}\\
    \midrule
    \multicolumn{3}{c}{\mylegend{Never}{blue3} \mylegend{Very rarely}{blue2} \mylegend{Rarely}{blue1} \mylegend{Occasionally}{orange1} \mylegend{Frequently}{orange2} \mylegend{Very frequently}{orange3}} \\
    \midrule
    Q8. How familiar are you with limitations and shortcomings of generative AI? & Ethics & \importancebarchart{0}{0.047}{0.363}{0.325}{0.210}{0.054}{0}{4.7\%}{5.4\%}\\
    \midrule
    \multicolumn{3}{c}{\mylegend{Not at all familiar}{blue2} \mylegend{Slightly familiar}{blue1} \mylegend{Somewhat familiar}{gray1} \mylegend{Very familiar}{orange1} \mylegend{Extremely familiar}{orange2}} \\
    \bottomrule
    \end{tabular}
\caption{AI literacy questions, their themes, scale, and distribution. $R$ denotes reversed scale.}
\label{tab:app-ai-literacy}
\end{table*}
\subsection{AI Literacy}
\label{app:ai-literacy}
In addition to demographics, participants were asked questions about their experiences with AI. 
More specifically, to understand participants' familiarity with AI, we asked six questions (Q1 through Q6) to assess AI awareness, usage, evaluation, and ethics from \citet{wang23literacy} with two additional questions to assess frequency of AI usage (usage) and familiarity with limitations of AI (ethics).
As shown in Table~\ref{app:ai-literacy}, majority of the participants were neutral or agreed with statements of familiarity with AI awareness, usage, evaluation, and ethics. 

\section{Coding Procedures}
\subsection{Pre-processing Details}
\label{app:parsing}
\paragraph{Prompt}
The prompt, system and user input, for parsing Q6 were the following:
\begin{itemize}
    \item[] \texttt{"system": "You are a helpful research assistant. You are processing public survey data for a project involving tasks that AI can be used for. Carefully read the question and participant answer, and parse the answer into separate tasks. All outputs should be in a json format."}
    \item[] \texttt{"user": "The following is the participant answer to the question "What are some tasks that you might have Tech-X 10 help with or automate? This is a brainstorming exercise! Feel free to answer with whatever comes to your mind."\textbackslash n\textbackslash n participant answer: "\{\}"\textbackslash n What are the mentioned tasks? Separate each task and provide them in a json format."}
\end{itemize}
For Q3, we used the same prompt with a different question, and the curly brackets (\{\}) denote where participant answers would be filled in. 
For both Q3 and Q6, two parsed examples were given as fewshot examples. 
\paragraph{Settings} 
We used \texttt{gpt-4-1106-preview} with the following settings: \texttt{max\_tokens=128, temperature=0.0, top\_p=1.0, frequency\_penalty=0.0, presence\_penalty=0.0, seed=42}.
Moreover, for those stopped due to length, we increased the max\_tokens parameter to 256. 

\subsection{Open Coding}
\label{app:coding}
The authors then performed open-coding on the 80 samples by inductively and independently generating codebooks. 
Codes were defined with a brief name and a description.
Each sample was labeled with one or more codes.
The authors then convened to merge the individual codebooks into a shared codebook by identifying similar codes and creating new a code for it.
For the remaining codes, the authors unanimously agreed to add or delete the code.
The shared codebook was then reapplied to the 80 data samples by unanimous vote by the three study authors.
Finally, the study authors reconvened to organize the codes around themes, which were decided based on a unanimous vote.

To improve the clarity of the codes' names and definitions, we applied GPT-4 on the 80 samples and reviewed areas of disagreement between the authors and GPT-4.
Only in the case where disagreements arose due to unclear or vague code definitions or names, the field was updated. 
This iteration, however, was only applicable for one of the questions (Q7).

This open coding process was utilized for data on use cases (Q7), harms and benefits of developing (Q8, Q11, Q14), harms and benefits of not developing (Q17, Q20). 
We applied codebook developed from open coding use cases for data on brainstormed answers (Q3, Q6).
For data on impacted groups by AI, as the answers were short and direct, a single author open coded 80 instances per question to develop the codebook and validation sets to measure agreement with GPT-4.

\subsection{Closed Coding Setting}
\label{app:closed-coding}
\paragraph{Prompt}
We used the following prompts to apply our codebook:
\begin{itemize}
    \item[] \texttt{
    "system": "You are a research assistant helping open coding of survey data. Carefully read the definition of each code and apply one or more codes to the participant's answer. All outputs should be in a json format."}
    \item[] \texttt{"user": "Following are the codes and their definition.\textbackslash n\{\}\textbackslash n\textbackslash nThis specific survey question asked: \"What is one task you think AI will change the most drastically?\" \textbackslash n\textbackslash nSelect *one to four* most relevant codes **from the codes defined above** for the following participant answer. Format output into a json.\textbackslash nparticipant answer: "\{\}"}
\end{itemize}
The above prompt was for coding Q7, where first curly bracket was filled with codes and definitions and the second filled with participant answers. 
For all questions requiring coding, we followed similar template with the questions changed to reflect the original question.
For all questions, two fewshot examples were given. 
\paragraph{Settings} 
We used \texttt{gpt-4-1106-preview} with the following settings: \texttt{max\_tokens=128, temperature=0.0, top\_p=1.0, frequency\_penalty=0.0, presence\_penalty=0.0, seed=42}.

\begin{table*}[htpb]
    \centering
    \footnotesize
    \begin{tabular}{l| c | c c c c c| c c c l l 
 }
    \toprule
        & Tasks  & \multicolumn{5}{c|}{Harms and Benefits} & \multicolumn{5}{c}{Groups} \\ 
        \midrule
        Metric & Q7 & Q8 & Q11 & Q14 & Q17 & Q20 & Q9 & Q12 & Q15 & Q18 & Q21\\
        \midrule
        Avg. & .59 & .51 & .67 & .68 & .64 & .59 & .78 & .86. & .88 & .86 & .82\\
        Scott's $\pi$ & .59 & .51 & .66 & .67 & .62 & .57 & .77 & .85. & .87 & .85 & .82\\
        Cohen's $\kappa$ & .59 & .51 & .66 & .67 & .62 & .58 & .77 & .85. & .87 & .85 & .82\\
    \bottomrule
    \end{tabular}
    \caption{Agreement metrics between human and GPT-4 showing all moderate to substantial agreement. We report average observed agreement, Scott's $\pi$~\cite{scott1955pi}, and Cohen's $\kappa$~\cite{cohen1960coefficient}.}
    \label{tab:irr}
\end{table*}
\subsection{Closed Coding Evaluation}
As shown in Table~\ref{tab:irr}, GPT-4 shows moderate to substantial agreement over all questions. 

\section{Extended Analysis of Use Cases}
\label{app:use-case}

\begin{table*}[h]
    \centering
    \footnotesize
    \resizebox{\textwidth}{!}{
    \begin{tabular}{p{0.04\textwidth}p{0.24\textwidth}p{0.24\textwidth}p{0.24\textwidth}M{0.04\textwidth}M{0.04\textwidth}}
    \toprule
        PID & Task Description (Q7) & Harms of Not Dev. (Q17) & Benefits of Not Dev. (Q20) & Conf. (Q24) & Agr. (Q25)\\
        \midrule
        P189 & \textit{``public transportation that helps the disabled.''} & \textit{``wouldn't have a way to make things easier disabled people''} & \textit{``human workers would keep their jobs''} & $-4$ & $-3$\\
        P32 & \textit{``driving cars for visually challenged and/or physically challenged people.''} & \textit{``Visually/physically challenged people will miss out on being more independent in the day-to-day activities.''} & \textit{``there would be no risk of malfunction during driving task.''} & $3$ & $4$\\
        \midrule
        P175 & \textit{``buying groceries because then it could...change peoples’ diets to be healthier''} & \textit{``people would have to spend a day each week to buy groceries''} & \textit{``people would be forced to go outside and interact in society to buy groceries''} & $-2$ & $-2$\\
        P10 & \textit{``helping me watch my diet and groceries on-hand... It would save me a LOT of time having to shop for groceries twice a week myself.''} & \textit{``It wouldn't be harmful. People would just go about making...grocery decisions like they do now. There is no negative impact...other than the simple lack of progress.''} & \textit{``People MIGHT start taking the initiative to be more knowledgeable and involved with their own diet and health goals instead of relying on an automated tool.''} & $3$ & $5$\\
        \bottomrule
    \end{tabular}}
    \caption{Case analysis of tasks for which participants indicated should not be developed (negative confidence and perceived agreement scores) compared to similar tasks that indicated otherwise. As their effects were most significant, harms and benefits of not developing are shown for comparison.}
    \label{tab:tension-examples}
\end{table*}
\vspace{-1em}
\subsection{Case Analysis of Conflicting Decisions}
Table~\ref{tab:tension-examples} show conflicting examples of use cases. For detailed discussion see Section~\ref{ssec:tension}.

\begin{table*}[!htpb]
    \centering
    \footnotesize
    \begin{tabular}{P{0.25\textwidth}p{0.03\textwidth}p{0.03\textwidth}p{0.59\textwidth}}
    \toprule
    \belowrulesepcolor{cellgray}
    \multicolumn{4}{l}{\cellcolor[rgb]{ .921,  .921, .921}\textbf{\emph{Personal life applications}}}\\
    \hline
    \texttt{Search} & Q3 & P245 & ``\textit{Finding specific recipes with specific ingredients}''\\
    \texttt{Feedback} & Q3 & P107 & ``\textit{Providing ideas on how to improve in certain hobbies}''\\
    & & P46 & ``\textit{Advising on how to discuss sexuality with family without angering them or turning the conversation into an argument}''\\
    \texttt{Simplification} & Q3 & P210 & ``\textit{Helping people understand complex information related to healthcare, such as doctor's forms/letters, health insurance forms, taxes, etc.}''\\
    \texttt{Efficient data analysis} & Q6 & P75 & ``\textit{Help with resource allocation that maximizes benefit for utilities and food}''\\
    \texttt{Writing assistance} & & P51 & ``\textit{Writing a CV/resume that tailors the job description}''\\
    \texttt{Health} & Q3 & P14 & ``\textit{Creating an exercise routine}''\\
    & Q6 & P87 & ``\textit{Replace doctor visits for non-life-threatening ailments}''\\
    \texttt{Mental health} & Q6 & P143 & ``\textit{Monitoring health data and providing personalized insights and recommendations for maintaining physical and mental health}''\\
    \texttt{Personal finance} & Q6 & P184 & ``\textit{Assist with budgeting and finances for people who struggle with budgeting}''\\
    \texttt{Personal life productivity} & Q3 & P26 & ``\textit{Provide a schedule to accomplish everything I want to get done today or this week}''\\ 
    \texttt{Accessibility marginalized}$^{*}$ & Q6 & P184 & ``\textit{Make technology more accessible to those with limited understanding or disabilities}''\\
    \bottomrule
    \end{tabular}
    \caption{Examples of personal life applications, items that were coded with \texttt{everyday life assistance}, \texttt{everyday task automation}, etc., along with their additional characteristic codes. Code condensed due to space marked with ($*$).}
    \label{tab:personal-life-apps}
\end{table*}
\subsection{AI Use Cases for Personal Life}
\label{app:ssec-use-case-personal-life}
Use cases for personal life were more common in participants' answers compared to ones impacting work and society, resulting in a wide array of application ideas to improve everyday life as shown in Table~\ref{tab:personal-life-apps}.
One commonly observed type of application was information seeking such as \texttt{search}, \texttt{feedback}, and \texttt{simplification} in a more personalized and ``\textit{specific}'' (P245) ways that current search engines cannot yet provide.
Participants also emphasized tools for not only synthesizing large amount of public data such as research and information but also from personal data to provide ``\textit{personalized insights}'' (P143).
Other applications were in assistance or automation of everyday tasks such as email writing, cooking, shopping, and repairs. 
Additionally, participants showed interest in using AI for improving physical and mental health, for better resource and time management, and for providing accessibility to all these personal life tasks for those who have difficulties. 
%

\subsection{AI Use Cases for Society}
\label{app:ssec-use-case-society}
As participants brainstormed use cases for a futuristic version of technology (Q6) and selected an application with the most drastic change (Q7), discussion of societal applications increased.
Some common AI support included finding new and creative solutions to societal issues such as ``\textit{Helping corporations get out of the boxed idea of the bottom line and become stewards to this planet...}'' (Q7, P13) highlighting environmental challenges and ``\textit{solving or coming up with new way of finding a solution to poverty and homeless}'' (Q7, 68) and ``\textit{Propose ways to make education more affordable for all}'' (Q6, P67) focusing on inequality and resource allocation challenges.
Applications targeting current issues were also mentioned including ``\textit{helping to eliminate false facts and rhetoric, often hateful, from social and mainstream media.}'' (Q7, P161) and ``\textit{mediation between countries at war}'' (Q6, P206).

\section{Extended Analysis of Harms and Benefits of Developing}
\begin{figure*}[htpb]
  \centering
  \subcaptionbox{Benefits and harms scale averaged by task theme.\label{subfig:theme-dev-hb-scale}}{\includegraphics[width=0.45\textwidth]{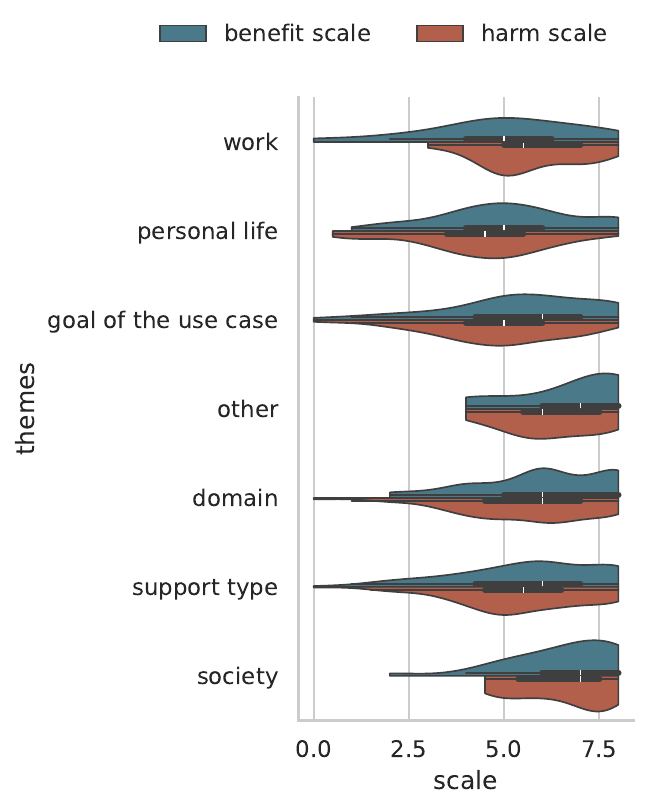}}%
  \subcaptionbox{Benefits and harms scale averaged by task \textsc{domain}.}{
    \includegraphics[width=0.45\textwidth]{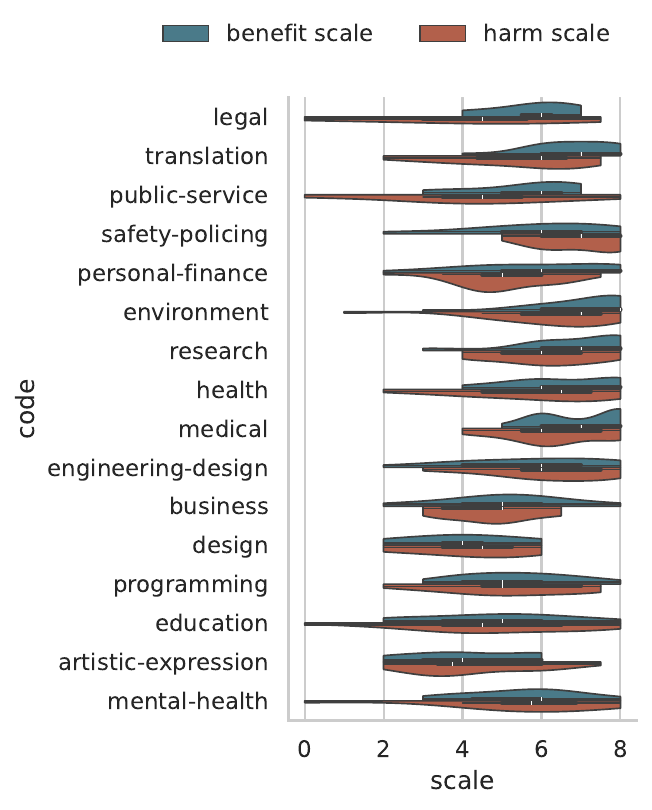}}
  \caption{Distribution of the harms and benefits scale by use case theme and domain sorted in order of decreasing absolute mean difference of benefit and harm. White mark indicates median, and black box within indicates quartile.}
  \label{fig:dev-hb-scale}
\end{figure*}

\subsection{Harms and Benefits of Use Cases}
\label{app:ssec-hb-usecase}
We plot the scales of impact for harms (Q13, Q16) and benefits (Q10) aggregated and averaged by themes and codes of corresponding use cases (Q7) in Figure~\ref{fig:dev-hb-scale}.
Noticeably, use cases that discuss \textsc{work} had a higher mean on the scale of harm (5.75$\pm$1.35) compared to benefit (4.94$\pm$2.08). 
\textsc{Personal life} use cases, on the other hand, had 
higher mean benefit (5.12$\pm$1.84) than harm (4.43$\pm$1.82). 
Disregarding the theme other, use cases that considers societal applications had the highest mean in both benefit (6.57$\pm$1.47) and harm (6.51$\pm$1.24).

Moreover, domains such as \texttt{legal}, \texttt{translation}, \texttt{public service} had the highest difference in their perception of benefit compared to their harms. 
While most domains had higher perceived benefit with AI applications compared to harms, domains such as \texttt{safety policing}, \texttt{engineering design}, and \texttt{design} had higher perceived harms compared to benefits. 
The domain of application that was perceived to be the most beneficial was in the use cases for the \texttt{environment} with mean of 7.00$\pm$1.41 on the scale of harm and safety policing was perceived to be the most harmful with a mean of 6.91$\pm$1.07 on the harms scale.

\begin{figure*}[htpb]
  \centering
     \includegraphics[width=\textwidth]{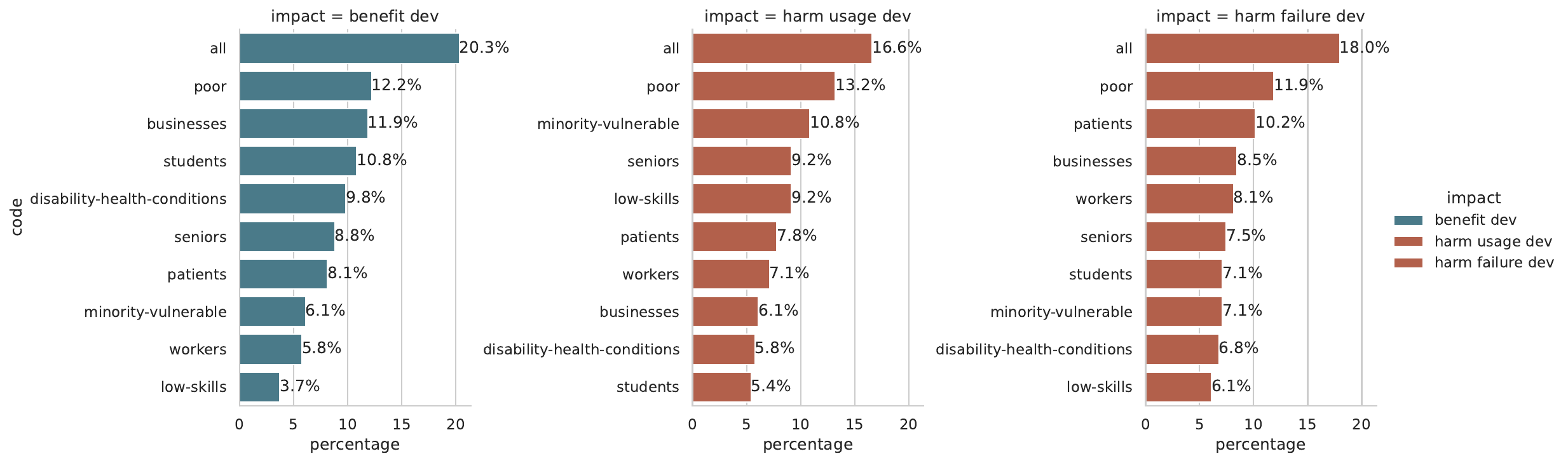}
  \caption{Distribution of top few \texttt{codes} mentioned in groups impacted the most by the use case (Q9, Q12, Q15).}
  \label{fig:groups-hb-dev}
\end{figure*}

\subsection{Groups Affected by Developing}
\label{app:ssec-groups-dev}


Some most frequently mentioned groups that participants selected to be benefiting or harmed the most by the use case are shown in Figure~\ref{fig:groups-hb-dev}.
The distribution of the top two most frequent were similar across the three questions, codes \texttt{all} (all people) (20.3\%; Q9, 16.6\%; Q11, 18.0\%; Q15) and \texttt{poor} (12.2\%; Q9; 13.2\%; Q11, 11.9\%; Q15), showing participants' interest in AI to improve accessibility and to help attain resources to improve the lives of everyone, especially those who do not have access due to lack of monetary means. 
However, the starting from the third most commonly affected groups, the distribution diverges.
\texttt{Businesses} were the third most commonly occurring group that would benefit the most. 
\texttt{minority and vulnerable} (10.8\%; Q12) were mentioned to be third most frequent as being harmed the most in cases of misuse, highlighting the understanding that AI applications might further drive inequality or would be misused to harm the vulnerable population.
\texttt{patients} (10.2\%; Q15) were also frequent in the groups to be harmed the most by failure cases, conveying the participants' interest in medical and health applications, however, failures being high risk.

\section{Extended Analysis of Harms and Benefits of \textit{Not} Developing}

\begin{figure*}[htpb]
  \centering
  \subcaptionbox{Benefits and harms scale of not developing averaged by task theme.\label{subfig:harms-themes}}{%
    \includegraphics[width=0.45\textwidth]{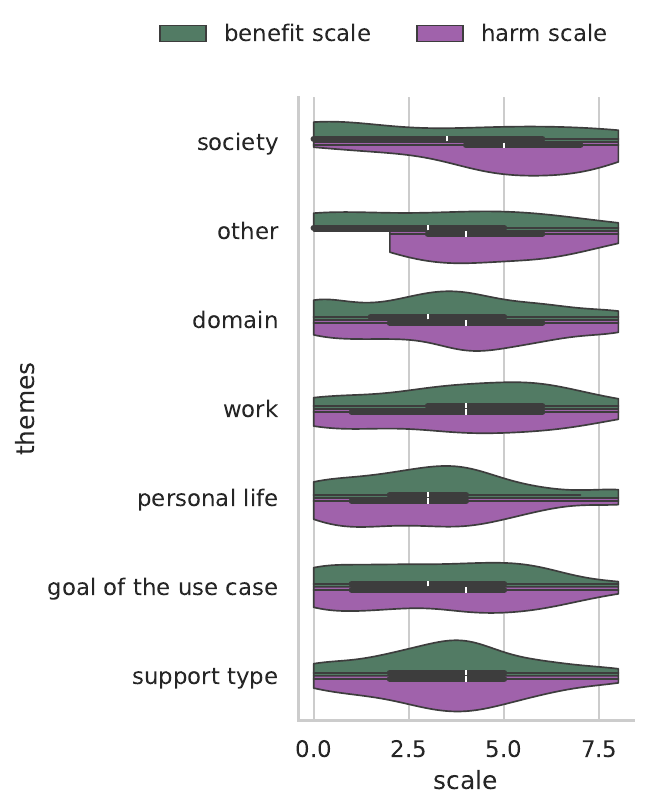}}
  \subcaptionbox{Benefits and harms scale of not developing averaged by task \textsc{domain}.\label{subfig:use-case-codes}}{%
    \includegraphics[width=0.45\textwidth]{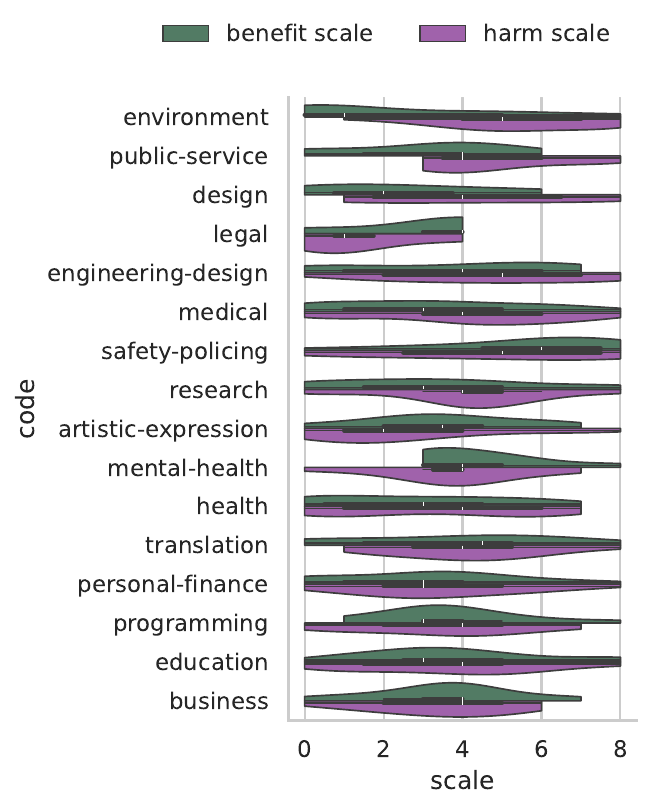}
  }
  \caption{not developing use case scale}
  \label{fig:not-dev-hb-scale}
\end{figure*}

\subsection{Harms and Benefits of Not Developing Use Cases}
\label{app:ssec-nd-hb-usecase}
Scale of harms (Q18) and benefits (Q21) of not developing aggregated by themes and codes of corresponding use cases (Q7) and sorted by descending order of mean absolute difference is shown in Figure~\ref{fig:not-dev-hb-scale}.
The use cases that impacted society had the highest mean difference between harms of not developing (5.14$\pm$2.12) and the benefits of not developing (3.32$\pm$3.15).
Work and personal life applications both had higher benefit of not developing, however, with personal life (3.32$\pm$2.27; Q21, 3.05$\pm$2.25; Q18) having a lower difference than work (4.11$\pm$2.38; Q21, 3.64$\pm$2.54; Q18). 
These results are consistent with the analysis detailed in Appendix~\ref{app:ssec-hb-usecase} of harms and benefits of developing in that work related applications are perceived to be more harmful to develop and beneficial to not develop, suggesting concerns of labor replacement. 
Moreover, uses of AI that helps societal issues are seen as having both high benefit and harms but also seen as harmful to not develop, indicating a fundamental tension.
\subsection{Use Cases: Should not Be Developed}
\label{app:ssec-usecase-should-not}
\begin{table*}
    \centering
    \footnotesize
\begin{tabular}{p{0.11\linewidth}p{0.42\linewidth}|p{0.40\linewidth}}
\toprule
\textbf{Theme} & \textbf{Example Tasks (Q7)}: \emph{Should not be Developed} & \textbf{Distribution} \\
\midrule
\textsc{Work} & \textit{``To replace employees in white collar jobs''} (P50) & \frequencybarchart{0.72}{0}{0}{0}{0.28}{0}{72\%}{28\%}\\
\textsc{Society} & \textit{``Assessing the metrics of a social problem...''} (P76) & \frequencybarchart{0.79}{0}{0}{0}{0.21}{0}{79\%}{21\%}\\
\textsc{Goal} & \textit{``public transportation that helps the disabled.''} (P189) & \frequencybarchart{0.82}{0}{0}{0}{0.18}{0}{82\%}{18\%}\\
\textsc{Domain} & \textit{``Give medical advice and health care prescriptions''} (P109) & \frequencybarchart{0.82}{0}{0}{0}{0.18}{0}{82\%}{18\%}\\
\textsc{Personal life} & \textit{``Create a meal plan and shopping list.''} (P242) & \frequencybarchart{0.86}{0}{0}{0}{0.14}{0}{86\%}{14\%}\\
\textsc{Support type} & \textit{``fact check political debates''} (P229) & \frequencybarchart{0.87}{0}{0}{0}{0.13}{0}{87\%}{13\%}\\
\textsc{Other} & \textit{N/A} & \frequencybarchart{0.92}{0}{0}{0}{0.08}{0}{92\%}{8\%}\\
\midrule
\multicolumn{3}{c}{\mylegend{Should be developed}{green3} \mylegend{Should not be developed}{red3}} \\
\bottomrule
\end{tabular}
\caption{Development opinions aggregated by use case (Q7) \textsc{theme}.}
\label{tab:task-theme-decision}
\end{table*}
Use case decision (Q23) aggregated by theme is shown in Table~\ref{tab:task-theme-decision}.
As discussed in the main results, \textsc{work} had the highest percentage of responses that the application ``Should not be developed'', and interestingly \textsc{personal life} application had the lowest percentage of the same answer compared to other realms of impact. 

\begin{figure*}[htpb]
  \centering
     \includegraphics[width=\textwidth]{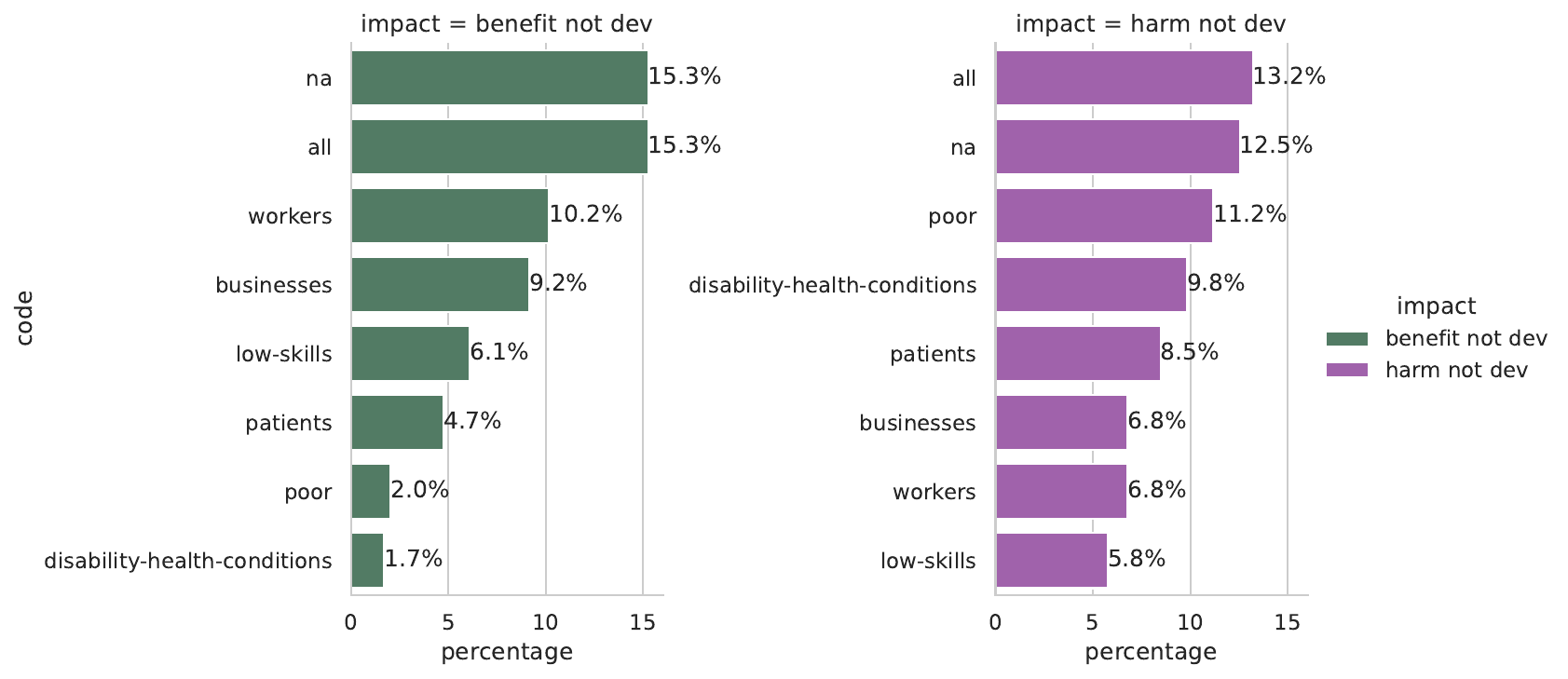}
  \caption{Distribution of top few \texttt{codes} mentioned in groups impacted the most by \textit{not} developing (Q18, Q21).}
  \label{fig:groups-hb-not-dev}
\end{figure*}

\subsubsection{Groups Affected by Not Developing}
\label{app:ssec-groups-nd}
Participants were less likely to write that any group would benefit or be harmed the most if the use cases were not developed (see Figure~\ref{fig:groups-hb-not-dev}). 
However, participants mentioned \texttt{workers} and \texttt{businesses} benefiting when the use cases were not developed more frequently compared to mentioning those characterized with fewer resources (e.g., \texttt{poor}) in contrast to those harmed from not developing. 
This again highlights the tension over AI development, which could help alleviate inequality of resources and can become harmful to workers.

\begin{table*}[htpb]
    \centering
    \footnotesize
    \begin{tabular}{l|lllllllll}
    \toprule
        Demographics & Age & Gender & Education & Political Leaning & AI Literacy Mean \\
        & ($N=295$) & ($N=280$) & ($N=287$) & ($N=295$)& ($N=295$) \\
        \midrule
        Decision & $n.s.$ & -0.197$^{*}$ & $n.s.$ & $n.s.$ & 0.238$^{**}$\\
        Decision$\times$Agreement & $n.s.$ & -0.218$^{*}$ & $n.s.$ & $n.s.$ & 0.272$^{**}$\\
        Decision$\times$Confidence & 0.189$^{*}$ & -0.265$^{***}$ & $n.s.$ & $n.s.$ & 0.297$^{**}$\\
        Benefits Dev. & $n.s.$ & $n.s.$ & $n.s.$ & $n.s.$ & $n.s.$\\
        Harms Dev. Misuse & 0.192$^{*}$ & $n.s.$ & $n.s.$ & $n.s.$ & $n.s.$\\
        Harms Dev. Failure & 0.216$^{*}$ & $n.s.$ & $n.s.$ & $n.s.$ & $n.s.$\\
        Benefits Not Dev. & $n.s.$ & $n.s.$ & $n.s.$ & $n.s.$ & $n.s.$\\
        Harms Not Dev. & 0.209$^{*}$ & $n.s.$ & $n.s.$ & $n.s.$ & $n.s.$\\
    \bottomrule
    \end{tabular}
    \caption{Correlations between worker demographics and use case decision, agreement, and confidence and their harms and benefits of developing and not developing. All variables were converted to integer scale. Variables that do not fall on a scale were converted as follows: Gender (M: 0, F: 1) and Political Leaning (Strongly liberal: 0, Strongly conservative: 4). AI Literacy Mean was calculated by taking the mean of AI literacy questions. Bonferroni corrected for multiple comparisons ($p < 0.0001: ***$, $p < 0.001: **$, $p < 0.01: *$).}
    \label{tab:app-demographics-analysis}
\end{table*}
\section{Extended Analysis of Demograhpic Factors}
To understand how participants of different demographics such as gender, age, and education level responded differently to the survey, we report statistical analysis results on participant responses based on their demographics. 
As shown in Table~\ref{tab:app-demographics-analysis}, AI literacy mean (Q1 through Q8 as shown in Table~\ref{tab:app-ai-literacy}) was positively correlated with decisions to develop the use case, perceived agreement of others to that decision, and confidence. 
Interestingly, our results showed that identifying as a female was negatively correlated with the decision to develop as well as perceived agreement and confidence and age was positively correlated with confidence and perceived harms of both developing and not developing.
The caveat is that all these analyses are confounded by the fact that participants wrote in their own use cases, and future work should study how user factors influence decisions controlling for use case.

\section{Extended Analysis of Conflicts of Harms and Benefits}
To qualitatively understand the conflicting values of harms and benefits, we ranked the harmonic mean (F1) of pairwise impact scales of four different combinations: Dev(BH), $\neg$Dev(BH), DevB$\neg$DevB, and DevH$\neg$DevH, where ``Dev'' and ``$\neg$Dev'' indicates developing and not developing the use case respectively, ``B'', benefit, and ``H'', harms.
The results shown in Table~\ref{tab:app-top-dilemmas}, indicate that use cases that target \texttt{medical}, \texttt{societal-issues}, and \texttt{education} could have higher perceived conflicting impact.

\onecolumn
{\footnotesize
\begin{longtable}{p{0.207\textwidth}p{0.2\textwidth}p{0.25\textwidth}p{0.25\textwidth}}
    \caption{Top 5 use cases ranked by harmonic mean (F1) of four different combinations of harms and benefits scales (Dev(BH), $\neg$Dev(BH), DevB$\neg$DevB, DevH$\neg$DevH). In case of ties, earlier submissions or non-repeating use cases were chosen. For Dev. Harms, ``(1)'' indicates harms of \textit{misuse} and ``(2)'' indicates harms of \textit{failure}, and the scale of impact for each types of harms were averaged.}
    \label{tab:app-top-dilemmas}\\
    \toprule
    \belowrulesepcolor{cellgray}
    \multicolumn{4}{l}{\cellcolor[rgb]{ .921,  .921, .921}\textbf{Top Dev. Harms and Benefits}}\\
    \hline
    Use Case & Codes & Dev. Benefits & Dev. Harms\\
    \hline
    \textit{``Developing medical care plans.''}& \texttt{medical}, \texttt{professional-consulting} & \textit{``It could focus on known medications, surgeries, and therapies that are assured to improve a patient’s health and health outcomes.''} & (1) \textit{``The outcome would be AI “deciding” who warrants health care/ which can led to death.''} (2) \textit{``The poor healthcare of people leading to perhaps death.''}\\
    \hline
    \textit{``I could go back to school and get a college degree, this could help me study.''} & \texttt{education}, \texttt{everyday-life-assistance} & \textit{``it can read text quickly and I am assuming turn any book into an audio book. If it has that capability I could go back to important chapters and paragraphs that are need to know information and listen to them, helping me remember better.''} & (1) \textit{``people could be poorly affected by misinformation''} (2) \textit{``students would get the wrong information and fail''}\\
    \hline
    \textit{``running simulations on future outcomes''} & \texttt{research}, \texttt{problem-solving} & \textit{``It would help to optimize present behaviors in pursuit of future success.''} & (1) \textit{``Severe income disparities, civil unrest, cultural wars, idol worship, public manipulation, and mass chaos''} (2) \textit{``Mass chaos and confusion throughout the world, regardless of socioeconomic or national identity''}\\
    \hline
    \textit{``How countries protect themselves against aggression and the threat of nuclear war.''} & \texttt{societal-issues}, \texttt{public-service} & \textit{``Countries would have time to alleviate the threat and use diplomacy.''} & (1) \textit{``Power is always the incentive for war. The negative impact would be possibly the total destruction of earth.''} (2) \textit{``Something or someone would be destroyed without notice.''}\\
    \hline
    \textit{``the medical field''} & \texttt{medical} & \textit{``it would lower cost of surgery's and procedures dramatically''} & (1) \textit{``It could cause world wars or devastation of the planet.''} (2) \textit{``Anyone who chose to allow the AI to perform medical procedures/surgeries.''}\\
    \midrule
    \belowrulesepcolor{cellgray}
    \multicolumn{4}{l}{\cellcolor[rgb]{ .921,  .921, .921}\textbf{Top $\neg$Dev. Harms and Benefits}}\\
    \hline
    Use Case & Codes & $\neg$Dev. Benefits & $\neg$Dev. Harms\\
    \hline
    \textit{``With the current economic recession in the world, how do i operate a healthy financial status?''} & \texttt{personal-finance}, \texttt{productivity}  & \textit{``People will be well informed and educated.''} & \textit{``People would have limited access to information''} \\
    \hline
    \textit{``Stop child trafficking''} & \texttt{societal-issues} & \textit{``To the traffickers''} &  \textit{``It will continue to happen''} \\
    \hline
    \textit{``How to solve climate change would be the most dramatic.''} & \texttt{research}, \texttt{problem-solving}, \texttt{societal-issues} & \textit{``it would not be used maliciously for the wrong reasons.''} & \textit{``humans would not be able to solve climate change without the technology, leading to catastrophic events.''}\\
    \hline
    \textit{``How countries protect themselves against aggression and the threat of nuclear war.''} & \texttt{societal-issues}, \texttt{public-service} & \textit{``Then we would not run the risk that AI would become its own entity and takeover its own programming''} &  \textit{``We would run the risk of not being fully prepared in case disaster was headed our way.''}\\
    \hline
    \textit{``The task that I think Tech-x 10 would most dramatically change will be helping with my house chores and taking care of my children and assisting them with their studies.''} & \texttt{companionship}, \texttt{education}, \texttt{everyday-task-automation} & \textit{``It will be beneficial because people won't be relying on any artificial intelligence to produce more ideas but would rather create more ideas''} & \textit{``It would be harmful because people will be stressed and depressed with plenty duties to attend to''} \\
    \midrule
    \belowrulesepcolor{cellgray}
    \multicolumn{4}{l}{\cellcolor[rgb]{ .921,  .921, .921}\textbf{Top Benefits}}\\
    \hline
    Use Case & Codes & Benefits Dev. & $\neg$Dev. Benefits \\
    \hline
    \textit{``Lifelong language learning: Master any language with a patient, AI tutor that adapts to your learning style and pace. Tech-X 10 can translate conversations in real-time, provide cultural context, and even help you practice your accent.''} & \texttt{companionship}, \texttt{education}, \texttt{translation}, \texttt{practical-skill-learning} & \textit{``Accessibility and affordability: Language learning would become accessible to everyone, regardless of location, socioeconomic background, or learning disabilities. Tech-X 10 would be a personalized and tireless tutor, eliminating the need for expensive private lessons or group classes.''} & \textit{``Focus on authentic communication: Without relying on AI translation, individuals would be forced to develop their own language skills, leading to a deeper understanding of grammar, vocabulary, and cultural subtleties.''} \\
    \hline
    \textit{``I'd have to say I think the most drastically changed thing would have to be the climate change issue because it's the most pressing to everybody's life, no matter where they live in the world right now. Nothing something with this kind of information and be able to think they can take in this much data in all that it can do could hit a drastic impact on the issues that we face as far as climate change is concerned.''} & \texttt{research}, \texttt{problem-solving}, \texttt{societal-issues}, \texttt{data-analysis} & \textit{``It would be beneficial because it could come up with answers to some of those pressing concerns about climate change areas where us humans are getting stuck on the we just may not have the correct answers or it can advance ideas and we already have that. We're just not sure how to implement and put in place.''} & \textit{``It's really hard to say, but again, say, for example, it did come up with a solution to 1 of the major problems of climate change. And then we did not implement. or develop. We would never have that beneficial answer that we could not come up with ourselves as humans. That could be catastrophic.''} \\
    \hline
    \textit{``the fields of science, education, and medicine. It would change how we learn, diagnose medical conditions, and develop technology.''} & \texttt{research}, \texttt{education}, \texttt{medical} & \textit{``The benefits would make education more accessible. It would save doctors time  and advance medicine. Doctors would have a wealth of information at their fingertips. It would also change the ways we develop technology and how we use technology.''} & \textit{``it would negatively impact those fields. The advancement in all of those fields would be delayed. It would take years instead of weeks or days using this technology. Tech-X / Tech-X 10 would save time.''} \\
    \hline
    \textit{``I think coding apps is the task that will be most drastically changed by Tech-X 10''} & \texttt{programming} & \textit{``It would help anyone make an app not just those who know how to code or have a lot of money. It would also make it much faster.''} & \textit{``Developers would be able to keep charging money for developing apps''} \\
    \hline
    \textit{``The task that I think Tech-X/Tech-X 10 would most dramatically change would be in personalized healthcare.''} & \texttt{medical} & \textit{``If Tech-X/Tech-X 10 was used to automate or assist with personalized healthcare, it would be beneficial because it could revolutionize medical recommendations and diagnostics, leading to more accurate treatments and empowering individuals to actively manage their health for improved well-being.''} & \textit{``If personalized healthcare was not automated or assisted by Tech-X/Tech-X 10, it could be beneficial by preserving the human touch in healthcare, maintaining the primacy of healthcare professionals' expertise, and ensuring a more personalized and empathetic patient-provider relationship, fostering trust, and potentially reducing reliance on technology for critical healthcare decisions.''} \\
    \midrule
    \belowrulesepcolor{cellgray}
    \multicolumn{4}{l}{\cellcolor[rgb]{ .921,  .921, .921}\textbf{Top Harms}}\\
    \hline
    Use Case & Codes & Harms Dev. & $\neg$Dev. Harms\\
    \hline
    \textit{``Medical diagnostics and treatment advancements''} & \texttt{medical}, \texttt{data-analysis} &  (1) \textit{``Patient harm and safety risks, Deterioration of healthcare trust, Public perception and resistance to adoption, Resource wastage, Security and privacy breaches''} (2) \textit{``Patient harm and worsened health outcomes, Loss of patient trust, Delay in proper medical care, Increased healthcare costs, Diminished adoption of AI in healthcare:''} & \textit{``it would pass up an opportunity to use new technology to improve healthcare accuracy, efficiency, and customization. This lack of automation may result in delayed diagnoses and treatment, decreased access to specialized knowledge, increased workload on healthcare professionals, limited personalization in treatment plans, increased risk of human error, missed opportunities for innovation, limited scalability of healthcare services, and less efficient use of resources. The capabilities of Tech-X 10 are critical in improving healthcare processes; without them, there is a high danger of detrimental effects on patient care and overall healthcare effectiveness.''} \\
    \hline
    \textit{``helping the police department.''} & \texttt{public-service} & (1) \textit{``citizens could be killed or attacked.''} (2) \textit{``It could kill the hostages or victims instead of apprehending the suspects.''} & \textit{``it would continue to kill innocent people.''} \\
    \hline
    \textit{``Advance robotics fields to assist with labor automation''} & \texttt{human-labor-replacement}, \texttt{physical}, \texttt{engineering-design} & (1) \textit{``Potential forced depopulation or enforced child limits.''} (2) \textit{``Civilization would be thrown back thousands of years in terms of progress.''} & \textit{``The current status quo would continue, where people's lives are being wasted on unfulfilling labor for low pay.''} \\
    \hline
    \textit{``Solving medical proplems would be a huge benefit to society as a whole.''} & \texttt{medical}, \texttt{problem-solving}, \texttt{societal-issues}  & (1) \textit{``People could be hurt or die.''} (2) \textit{``More people would get sick or die.''} & \textit{``New technology would not be used to help man kind.''} \\
    \hline
    \textit{``How to convert ocean water into drinking water.''} & \texttt{research}, \texttt{problem-solving}, \texttt{engineering-design} & (1) \textit{``People would literally be dehydrated and die for no good reason.''} (2) \textit{``Absolute chaos and the death of people unnecessarily.''} & \textit{``It would be against mankind to do what is best for people in the name of society.''} \\
    \bottomrule
\end{longtable}}
\twocolumn

\section{Codes}
\label{app:codes}
Here we detail codes developed in \S\ref{sec:method} in order as they appear in the main text.

\onecolumn
{\footnotesize
\begin{longtable}{p{3cm}|p{11cm}}
\caption{Codes and their definition for Q7 (tasks).} \label{tab:codebook-q7}\\
\toprule
Code & Definition \\
\hline
  education &  Applications for educational purposes that is traditionally tought in schools or follows a curriculum \\
 legal &  Applications in legal domain \\
 medical &  Applications in the medical domain \\
 health &  Applications in health and well-being \\
 mental-health &  Applications in mental health both specialized and therapeutic everyday assistance \\
 business &  Applications in businesses such as profit enhancement tools for growth projection supply chain applications advertisement etc. \\
 personal-finance &  Applications in personal finances such as bill paying insurance taxes etc. \\
 artistic-expression &  Applications in the arts such as story creation image generation for paintings etc. \\
 engineering-design &  Applications for creating engineering designs for houses cars buildings etc.  \\
 programming &  Applications for programming software applications \\
 public-service &  Applications for government work and public service \\
 physical &  Applications that offer physical assistance \\
 translation &  Applications for translation \\
 companionship &  Applications to function as a companion (e.g. babysitter bot to talk to when lonely tutor etc.) \\
 interpersonal-communcation &  Applications for assisting communication especially interpersonal (e.g. improve expression of self emotional / social connection etc.) \\
 practical-skill-learning &  Applications for practical skill learning such as repair cooking etc. \\
 everyday-task-automation &  Applications to automate mundane everyday tasks such as shopping meal-planning paying bill etc. \\
 everyday-life-assistance &  Applications to assist and optimize everyday life (e.g. organizing journal entries for self-discovery optimal scheduling) \\
 feedback &  Applications that can give suggestions and feedback for improvement \\
 fact-checking &  Applications to help with fact checking information \\
 workplace-productivity &  Applications to assist at workplace to increase productivity and hep with more mundane tasks but not fully automating the job \\
 human-labor-replacement &  Applications that replace human laborers such as robots robot servers which are more physical / menial but also experts such as legal medical financial business etc. \\
 professional-consulting-service &  Applications to give expert advice suggestions and services (e.g. medical or financial consulting) \\
 beyond-human &  Applications that leverages AI for beyond human capabilities \\
 efficient-data-analysis &  Applications that perform large scale data analysis or fast data analysis in a manner that is resource intensive for human alone - explicitly mentions data and analysis \\
 search &  Applications for search and sense making \\
 writing-assistance &  Applications for writing such as providing grammar edits or suggestions \\
 image-generation &  Applications for image generation from artistic images to practical ones such as concept art figures etc. \\
 creativity &  Application descriptions that specify a sense of creativity by using specific words that indicate creativity for example creative story telling creative advertisement etc. \\
 simplification &  Applications that simplify the task for example through summarization \\
 embodiment &  Applications that are embodied or require control in physical environment \\
 research &  Application that conducts open-ended research and performs knowledge discovery by asking and answering questions \\
 design &  Applications that aids design of objects and systems with a focus on creativity holistic approach user experience and aesthetics \\
 math-problem-solving &  The application is for solving math problems \\
 mystery-crime-solving &  The application is for solving mysteries and crimes \\
 safety-policing &  Applications for safety and policing purposes to reduce crimes \\
 environment &  Applications for environmental purposes such as reducing waste helping climate change etc \\
 societal-issues &  Applications for solving societal issues such as human trafficking cliamate change etc. \\
 brainstorming &  The answer mentions that the application's goal is to provide a new angle to solve problems using brainstorming \\
 business-productivity &  The answer mentions that the goal of the application is to help businesses to cut down on resources or create more output and profit \\
 personal-life-productivity &  The answer mentions that the goal of the application is to make them more productive in their life personally or to save a lot of time for themselves \\
 societal-productivity &  The answer mentions that the task will be more efficient and the application will help make the society more productive \\
 accessibility-marginalized-disabled &  The participant answer mentions that the goal of the application is to marginalized and disabled people will be able to get help \\
 lower-barriers-resources &  The participant answer mentions that the goal of the application is to lower barrier to resources \\
 new-code &  None of the above codes apply but the answer is still meaningful so a new code is needed \\
 na &  The participant answer does not make sense in the context \\
 \bottomrule
\end{longtable}}
\twocolumn
\onecolumn
{\footnotesize
\begin{longtable}{p{3cm}|p{11cm}}
\caption{Codes and their definition for Q11 \& Q14 (harms of developing).} \label{tab:codebook-q11-q14}\\
\toprule
Code & Definition \\
\hline
  manipulate-people & Harm that misleads people to make choices that do not benefit them and is deceptive or fraudulent\\
misinformation & Harm that causes people to believe in false information or incorrect state of the world by intentionally providing misleading knowledge i.e. spreading misinformation\\
bias & Infringement of social justice by spreading prejudice and bias\\
mental-harm & Mentally harm or upsets people by hurtful outputs and spread of negative information\\
overreliance & People becoming dependent on technology and overtrusting and overrelying on them leading to diminished abilities to complete the task\\
physical-harm & Technology leads to physical harm such as injuries and deaths\\
war & Technology is used for wars or leads to wars and physical harm at a societal or global level\\
economic-disturbance & Technology causing wider economic harms and disturbances such as widespread job loss or depression\\
financial-disturbance & Technology causing more individual or smaller scale financial loss or property damage\\
human-labor-replacement & Technology causes job loss and replacement of human labor force causing unemployment\\
social-isolation & Technology causes weakened interpersonal connection especiallh with family and friends leading to isolation\\
range & Technology causes a range of harms from very small impact to serious and more wide-spread harms\\
aid-criminal & Technology is used to aid criminal activity\\
distrust-ai & Technology or complicated output leads to distrust or underuse of the AI application\\
distrust-institution & Technology leads to distrust of institutions such as the healthcare system\\
data-security-privacy-risk & Privacy is invaded or data is lost through the use of technology or data is used in a negative way to benefit other stakeholders rather than the user\\
plagirism & Technology plagirizes the existing work or copyrighted work\\
damaging-creativity & Technology damages creativity or leads to unoriginality\\
hinder-career & Technology causes career damage\\
incorrect-ai-output & AI output being unintentionally incorrect or erroneous leads to different harms to users such as misdiagnosis or incorrect advice\\
legal-issues & AI causing legal issues such as law suits due to illegal outputs\\
impede-learning & AI causes people to not learn or grow as much\\
social-division & AI causes social division and leads societal spread of hate or distrust\\
extinction & AI leads to extinction of some sort such as group of people human race other animals or culture\\
minority & AI leads to harming minority or underrepresente groups\\
waste-resources-or-time & Technology leads to wasting resources such as time in development and is useless\\
unqualified-accessibility & Technology makes certain tasks too easy so that non-qualified people or bad actors have better accessibility to these tasks\\
terrorism & Technology leads to aiding terrorists for example in making weapons or bombs\\
business-use & Technology is used by businesses to maximize profit\\
non-war-military-use & Technology is used by the military for non-war purposes\\
lower-quality & Technology lowers the quality by making mistakes or creating homogenious outputs which are worse than human work\\
information-access & Technology prevents access to information\\
general-harm & Risks security through lack of safety checks or the application is rendered unsafe and can harm or hurt users in unspecified ways\\
hinder-science & Hinders scientific breakthroughs\\
no-harm & No harms caused\\
miscommunication & The use of application leads to miscommunication\\
negative-health-wellbeing & Technology causes negative health outcome\\
hinder-medical-care & Technology causes hindrance to medical and healthcare advancement and application\\
environmental-harm & Causes environmental harm or allows continued environmental harm such as climate change\\
hacking-risk & AI could be hacked by bad actors to be used for malicious tasks\\
new-code & None of the above codes apply but the answer is still meaningful so a new code is needed\\
na & The participant answer does not make sense in the context\\
 \bottomrule
\end{longtable}}
\twocolumn
\onecolumn
{\footnotesize
\begin{longtable}{p{3cm}|p{11cm}}
\caption{Codes and their definition for Q8 (benefits of developing).} \label{tab:codebook-q8}\\
\toprule
Code & Definition \\
\hline
scientific-research-innovation & Advances science research innovation and discovery\\
improve-medical-care & Improves medical care\\
specialized-resources-accessibility & Allows specialized resources such as medical mental legal or financial resources and services more available\\
information-accessibility & Allows information to be more accessible\\
resource-accessibility & Provides resources more accessible for everyday or life tasks\\
personal-life-efficiency & Allows people to be more productive with less time and effort to solve tasks faster and easier\\
reduce-mundane-work & Reduces mundane work in everyday life\\
improve-mental-health & AI application improves mental health\\
companionship & Provides companionship\\
social-interaction & Provides assistance in social interaction\\
better-communication & Allows people to communicate more effectively especially with less misunderstanding\\
more-production & Application allows more production of goods or services through automation or reducing overhead of human labor\\
personal-growth & Allows opportunities for personal growth or learning\\
enhancing-creativity & AI tool inspires more creativity\\
financial-gain & AI application leads to financial gain\\
safety & AI helps make the world safer by handling dangerous situations or improving policing\\
less-human-error & AI helps reduce human errors or bias\\
information-quality & AI assists in improving information quality by fact-checking or ensuring that the infromation is correct\\
improve-well-being-health & AI improves well-being fitness and health\\
improve-societal-issues & AI improves societal issue\\
save-life & The tecnology can save lives\\
general-efficiency & The technology offers general efficiency in speeding up the process or reducing needed resources\\
no-benefit & The answer says there is no benefit to the technology\\
new-code & None of the above codes apply but the answer is still meaningful so a new code is needed\\
na & The participant answer does not make sense in the context\\
 \bottomrule
\end{longtable}}
\twocolumn
\onecolumn
{\footnotesize
\begin{longtable}{p{3cm}|p{11cm}}
\caption{Codes and their definition for Q17 (harms of not developing).} \label{tab:codebook-q17}\\
\toprule
Code & Definition \\
\hline
less-innovation & Not using AI will lead to less innovation leading to stagnation of society\\
    delay-in-innovation & Not using AI will delay the rate of growth or breakthroughs however it might be possible for humans to get there without AI but just slower\\
    misinformation & Lack of AI usage continues division or mislead people with incorrect information\\
    waste-resources-or-time & Not using AI makes the task less efficient\\
    business-loss & Businesses or companies losing profit\\
    financial-disturbance & Financial loss at a smaller scale such as personal finance\\
    unemployment & People losing jobs\\
    physical-harm & Not using AI leads to physical harms such as death and injury\\
    impede-personal-growth & People lose the opportunity to grow or achieve without the help of AI\\
    human-error & Without AI human errors can be harmful\\
    mental-harm & Not developing the application leads to worse mental health such as anxiety depression loneliness etc. \\
    health-issues & Lack of AI assistance causes people to be unhealthy\\
    stress-overworked & Causing people to be overworked or be stressed because of the lack of automation offered by AI\\
    inefficiency & People will have to find another way that is not dependent on AI to solve the issue would cause some inconvenience but not disruptive\\
    environmental-harm-continues & Global warming and other environmental issue continues\\
    lose-transparency & Lack of AI assistance to understand complex systems leads to less transaprency\\
    hinder-communication & Without the application there will continue to be misunderstandings and difficulties in communication\\
    hinder-creative-work & Without the application products will become less creative limited to human creativity\\
    lose-tech-race & Lack of development will lead to losing technical race between countries and cause political tension\\
    hinder-medical-care & Not devloping the application will hinder patients from getting better medical care or treatment\\
    lose-information-knowledge & Not developing the application will lead to loss of information and knowledge\\
    lose-accessibility-solution-service & Not developing the application will lead to losing one of the solutions to a problem or service\\
    lose-assistance & Not developing the application will result in less help and assistance for the task\\
    economic-distrubance & Leads to economic distrubance such as cost increases at a larger scale\\
    no-harm & There is no harm of not developing the application\\
    new-code & None of the above codes apply but the answer is still meaningful so a new code is needed\\
    na & The participant answer does not make sense in the context\\
 \bottomrule
\end{longtable}}
\twocolumn
\onecolumn
{\footnotesize
\begin{longtable}{p{3cm}|p{11cm}}
\caption{Codes and their definition for Q20 (benefits of not developing).} \label{tab:codebook-q20}\\
\toprule
Code & Definition \\
\hline
less-dependent-on-tech & Make people less dependent on technology and self reliant in that they will have the skills to complete the tasks themselves\\
less-improper-unethical-use & Generally reduces misuse or ethical concerns of AI\\
relieve-plagarism & Relieves plagiarism issues\\
more-privacy & Preserves data privacy\\
job-security & Preserves employment and job security\\
learning-skills-knowledge & People would learn more without AI\\
human-interaction-dependence & People will interact more with real people and not AI increasing social interaction and interpersonal relations learning to depend on each other and invest in each other\\
environmental & The environmental harms will be reduced\\
creativity & Without AI people will be more creative and outputs will be more unique\\
less-misinformation & Using AI for generation or spreading of misinformation would be avoided and reduced\\
maintain-status-quo & Without the disruption of AI the current world will continue as is i.e. social order will not be disrupted and will continue to develop at the current pace\\
financial-benefit & Without AI people will continue to pay for services and their providers as before increasing their financial benefit\\
empathy & Without AI interactions and services will be more empathetic\\
higher-quality & Humans will create higher quality outputs with less ai errors\\
better-health & Results in better physical and mental health\\
human-control & Humans will be able to control from their understanding of the process for certain tasks\\
more-attentive & People will be more attentive to the task and lead to more understanding of the underlying problem\\
human-brilliance & The world will rely more on human brilliance leading to more investment and celerbration of human ingenuity\\
other-non-ai-solutions & Development will happen even if the application is banned through other non-ai solutions\\
no-benefit & Participant answer specifies that there is no benefit\\
new-code & None of the above codes apply but the answer is still meaningful so a new code is needed\\
na & The participant answer does not make sense in the context\\
 \bottomrule
\end{longtable}}
\twocolumn
\onecolumn
{\footnotesize
\begin{longtable}{p{3cm}|p{11cm}}
\caption{Codes and their definition for coding groups (Q9, Q12, Q15, Q18, Q21).} \label{tab:codebook-groups}\\
\toprule
Code & Definition \\
\hline
researchers-scholars & People who are researchers (e.g. scientists) or scholars experts\\
low-skills & People with limited skills education knowledge or critical-thinking\\
adults & People who are adults but not elderly\\
youth & Young people\\
seniors & People who are old\\
lawyers & People who are lawyers\\
lawyer-clients & People who are clients of lawyers\\
rich & People who are rich or in a high socioeconomic status\\
poor & People who are not rich or engage in risky financial habits\\
middle-class & People who are in the middle class\\
tech-access & People who do have access to cutting-edge technology\\
no-tech-access & People who do not have access to cutting-edge technology\\
it-professionals & People who work in the IT industry such as software engineers\\
engineers & People who are engineers\\
internet-users & People who use the internet\\
anti-technology & People who are skeptical of technology\\
teachers & People who teach others\\
students & People who are students or engage in learning\\
coaches & People who are coaches\\
athletes & People who are athletes\\
english-speakers & People who only speak English for a language\\
nonenglish-speakers & People who do not speak English\\
citizens & People who are American citizens and have privileges only granted to these citizens such as voting\\
immigrants & People who are immigrants\\
developing-nations & People who live in developing nations\\
travelers & People who travel or are interested in other cultures\\
democrats & People who are Democrats or left-leaning\\
republicans & People who are Republicans\\
doctors-nurses & Healthcare professionals such as doctors and nurses\\
patients & People who are patients or are receiving healthcare\\
businesses & Businesses people who own businesses or high-level executives\\
consultants & People who are consultants\\
consumers-stakeholders & People who are consumers or stakeholders of a service or product\\
workers & People who work professionally\\
busy-people & People who are busy or have limited time due to other pressing commitments\\
disability-health-conditions & People who have physical and mental disbilities or preexisting or chronic health conditions\\
nd-people & People who are neurodivergent\\
mental-health & People who have mental health conditions\\
single & People who are not married or single\\
families & A family\\
religious-minority & People who belong to a minority religious gruop\\
christian & People who are Christian\\
racial-minority & People who are racial minorities and are not White\\
white & People who are White\\
do-drive & People who drive\\
don't-drive & People who do not drive or are passengers\\
men & People who are men\\
gender-minority & People who belong to a minority gender group\\
lgbtq & People who belong to the LGBTQ community\\
government-officials & Government institutions or people who work for such institutions\\
crime-victims & People who are negatively impacted by a crime including crime victims victims' families or people falsely accused of crimes\\
activists & People who advocate for any social cause\\
criminals & People who are criminals or engage in illegal activity\\
all & All people in the world\\
minority-vulnerable & People who belong to groups that are minoritized or generally vulnerable\\
na & There is no valid group\\
remote-location & People who live in remote areas\\
urban & People who live in urban environments\\
artists-creatives & People who are artists or engage in creative work\\
chefs & People who cook\\
high-power & People who have large amounts of influence or power\\
me & The response references the person writing the response\\
 \bottomrule
\end{longtable}}
\twocolumn

\end{document}